\documentclass[aps,prd,onecolumn,superscriptaddress,nofootinbib]{revtex4-2}
\usepackage[utf8]{inputenc}
\usepackage{amsbsy}
\usepackage{amsmath}
\usepackage{amsfonts}
\usepackage{xcolor}
\usepackage{physics}
\usepackage{graphicx}
\usepackage{mathrsfs}
\usepackage{xcolor}
\RequirePackage[colorlinks,hyperindex]{hyperref}
\hypersetup{colorlinks=true,breaklinks=true,urlcolor=blue,linkcolor=red}

\begin{document}
\title{Quantized Dirac Fields in torsionful gravity: cosmological implications and links with the dark universe}
\author{Antonio Capolupo}
\email{capolupo@sa.infn.it}
\affiliation{Dipartimento di Fisica ``E.R. Caianiello'' Universit\'a degli Studi di Salerno,
  and INFN   Gruppo Collegato di Salerno, Via Giovanni Paolo II, 132, 84084 Fisciano (SA), Italy}

\author{Sante Carloni}
\email{sante.carloni@unige.it}
\affiliation{DIME Sezione Metodi e Modelli Matematici, Universita di Genova, Via all'Opera Pia 15, 16145 Genova, Italy}

\author{Luca Fabbri}
\email{luca.fabbri@unige.it}
\affiliation{DIME Sezione Metodi e Modelli Matematici, Universita di Genova, Via all'Opera Pia 15, 16145 Genova, Italy}

\author{Simone Monda}
\email{smonda@unisa.it}
\affiliation{Dipartimento di Fisica ``E.R. Caianiello'' Universit\'a degli Studi  di Salerno,
and INFN   Gruppo Collegato di Salerno, Via Giovanni Paolo II, 132, 84084 Fisciano (SA), Italy}

\author{Aniello Quaranta}
\email{anquaranta@unisa.it}\affiliation{School of Science and Technology, University of Camerino, Via Madonna delle Carceri,	Camerino, 62032, Italy}

\author{Stefano Vignolo}
\email{stefano.vignolo@unige.it}
\affiliation{DIME Sezione Metodi e Modelli Matematici, Universita di Genova, Via all'Opera Pia 15, 16145 Genova, Italy}
\date{\today}
\begin{abstract}
We consider a classical field in square torsion theory as a source of torsion for a quantum fermion field in FLRW metric. In the framework of QFT, we obtain vacuum contributions to the energy-momentum tensor and to the axial current that modify the dynamics of the classical field and the field equations as back-reaction. These contributions lead to a modified classical field and therefore to a modified  torsion term $L^\mu$ and expectation value of energy-momentum tensor $T^{\mu\nu}$ on the quantum vacuum, altering the field equations in an interative process. We consider the first step of this process and we find that the vacuum condensate could affect the inflationary phase of the Universe. Higher order terms could impact the dark Universe.
\end{abstract}
\maketitle
\section{Introduction}

With the detection of gravitational waves \cite{Ligo0,Ligo1,Virgo0,Virgo1,Virgo2}, another experimental prediction of Einstein gravity has now been observed. In fact, if dark matter \cite{Rubin1,Rubin2,Rubin3,Rubin4,Rubin5} is indeed a form of matter \cite{PecceiQuinn77,DMAssion0,DMAssion1,DMAssion2,DMAssion3,DMAssion4,DMAssion5,Axion4}, and not a gravitational effect, there are at the moment no observational issues left open in modern gravity.

Nevertheless, from a purely theoretical perspective, there is a problem that needs fixing, and that is, the nature of singularities, arising in regions of infinite density that appear in Black Holes and at the Big Bang. Physically, their occurrence is ensured by the fact that gravity is always attractive, so that matter distributions tend to increase in density, in turn increasing the gravitational pull even more, in an unavoidable collapse. Mathematically, their presence is guaranteed by the Hawking-Penrose (HP) theorem \cite{HP1969} together with specific energy conditions.

One way out is to circumvent the HP theorem altogether by working in some theory of gravity that is not Einsteinian by considering higher-order derivative field equations \cite{DerivativeOrder0,Capozziello2011,DerivativeOrder1,DerivativeOrder2,DerivativeOrder3}. Alternative Einstein gravitational theories are widely taken into account at the moment, with the $f(R)$ types being just the most famous to have arisen in recent times (for a general over-view, we refer the reader to \cite{Cai:2013lqa, Nashed:2020kdb, Capozziello:2011gw, Vignolo:2018eco, Vignolo:2019qcg} and references therein).

Within Einstein gravity, where the HP theorem is enforced, the other way out is to avoid the hypotheses of the theorem in the first place. This means finding extensions of Einstein gravity where the energy tensor is modified, and thus the energy condition is violated. When considering the mathematical setting of Einstein gravity, that is Riemann geometry, there is one extension that appears to be natural. It is the Riemann-Cartan geometry, where the connection can be written as the sum of the Levi-Civita connection plus contributions due to the torsion tensor $T$  \cite{Torsion1,Hehl1976,Torsion2,Torsion4,Torsion5,Torsion6, DETorsion, Cirillo-Lombardo, FV1, VFC}. When the Riemann-Cartan geometry is taken as basis for the physical theory, the result is the Einstein-Sciama-Kibble gravity with torsion, where the curvature tensor still couples to the energy, according to the Einstein--like equations, but in which torsion couples to the spin, according to the Sciama-Kibble equations \cite{S,K,Kibble1967,Kibble1976,Kibble1980,Kibble1982,KibbleZurek1983}. The theory is therefore equipped to host any form of matter having both energy and spin, which means spinor fields. When spinor fields are present, their coupling to torsion induces a modifications of the energy tensor, carrying extra terms in the energy conditions, which then are expected to be violated.

The initial hope, nevertheless, was soon to be abandoned since it was proved that these extra terms actually worsen the conditions for which singularities can form \cite{k, Inomata:1976wi}. However, not all is lost. Indeed, the Sciama-Kibble completion of Einstein gravity, while being an extension of the original theory, it is only the simplest of all possible extensions. More in detail, Einstein gravity is based on the field equations obtained by varying the Hilbert Lagrangian $\mathscr{L}_{\mathrm{E}}\!=\!R(g)$ where $R(g)$ is the Ricci scalar of the Levi-Civita connection entirely written in terms of the metric. Its Sciama-Kibble completion is based on field equations obtained by varying the Lagrangian $\mathscr{L}_{\mathrm{ESK}}\!=\!R(g,T)$ where now $R(g,T)$ is the Ricci scalar of the full connection accounting for both the metric and the torsional contribution. This Lagrangian can be explicitly written as $\mathscr{L}_{\mathrm{ESK}}\!=\!R(g)\!+\!T^{2}/4$ in which $T^{2}$ is a specific square torsion term. But because torsion is a tensor, it is possible to add further square torsional contributions shifting the Lagrangian to a more general $\mathscr{L}^{\mathrm{full}}_{\mathrm{ESK}}\!=\!R(g)\!+\!kT^{2}$ whose field equations, in presence of spinors, would also modify the energy tensor. As the energy condition is modified with terms depending on the constant $k$, which is undetermined, they could be violated, if the sign of $k$ is chosen wisely \cite{Magueijo:2012ug, Khriplovich:2013tqa, Alexander:2014eva, Buchbinder:1985ym, Fabbri:2012yg, VFS}.

The avoidance of singularities in the case of the Big Bang, may be especially relevant in cosmological contexts about the issue of Big Bounce universes. Several studies have explored various aspects of quantum physics in curved spaces \cite{Shapiro,QCD,QCD2,Casimir,Hawking,Unruh,Takagi,Vanzella,Parker,Parker2}.
Here, we focus on the square-torsion modification of the Einstein-Sciama-Kibble gravity: in this case, torsion is sourced by a classical Dirac field $\phi$. In a spatially flat FRW universe, the resulting coupled Dirac equations can be solved, leading to an exact expression for torsion in terms of the scale factor. On such a background we study a quantized Dirac field $\psi$, in order to go beyond the classical result onto a semiclassical approach, where the quantum field affects the classical gravitational dynamics through expectation values.

It can be expected, when quantum fields are introduced in the semiclassical framework, that significant modifications to the gravitational dynamics, and therefore to the resulting cosmology, may appear \cite{Lepto1,Kaplan}. This is the case for instance for mixed quantum fields, which have been shown to give rise to a possible dark-matter like compononent\cite{N1,N2,N3,N4,N5-0,N5,N5-1,N6,N7,N8,N9,N10,N11,N12,N13,N14,N15,N16,N17, Torsioncostant,Neut2,Neut4,Neut5,Neut6,K4, K2, K5, Barker1978, NTorsion7, SUGRA, Jenkins, Sakharov}. The emergence of quantum sources is accompanied with a nontrivial condensate structure of the vacuum state for the quantum field, as a consequence of the gravitational background \cite{Pisacane}. Condensates are crucial for understanding the cosmos \cite{CapDark1, CapDark2,CapDark3, CapDark4, CapDark5,GPL,Cabral,Cap2,Capoz1,Propagating1,Propagating2,Propagating3,Cosm1, LSB, LSB2}. 
In the present case, the Hamiltonian for the quantum fermion field $\psi$ is not diagonal, due to the torsion. We diagonalize it using a Bogoliubov transformation, which leads to a new Fock space representation \cite{Campbell, Campbell2, Campbell3, Campbell4}. This representation is unitarily inequivalent to the Fock space of the quantum field $\psi$ evolving in the absence of torsion. As a result, the vacuum state derived from this transformation, denoted as $| 0_c(t_0) \rangle$, is orthogonal to the original vacuum $|0\rangle$ in the infinite volume limit \cite{Umezawa1,Umezawa2}. In particular,   the vacuum $| 0_c(t) \rangle$ acquires a condensate structure which induces to contributes of the vacuum expectation values of the energy-momentum tensor and of the axial current 
(as also discussed in \cite{Quaranta-Capolupo}).

These contributions act as additional sources for the torsion field. Consequently, these new terms modify the dynamics of the classical field $\phi$, affecting the field equations through back-reaction. To incorporate these effects,  we have to consider a new classical field with a new torsion term, which generates additional contributions to the vacuum expectation values of the energy momentum tensor  and of  the axial current, thus conditioning the field equations in an iterative process. We study the  first step of this iteration.  We find a contribution to both the energy and torsion that scales as $C^{-4}(t)$ in the strong coupling limit, implying possible influences of
   the condensate structure of the vacuum on the inflationary phase of the Universe.

The analysis of the full back-reaction mechanism will be analyzed in a further coming paper.
We expect that the vacuum terms derived considering also other steps of the iterative process,  could affect the dark sector of the Universe.

The paper is structured as follows. In section II we present the classical square torsion theory, solving the Dirac Equation for $\phi$ and determining the corresponding torsion term. In section III we introduce the quantum Dirac field $\psi$ on the previously described background. We solve the Dirac equation exactly and, diagonalizing the field Hamiltonian, we determine the physical vacuum. In section IV we compute the expectation value of the spin density on the physical vacuum, and we show that it is non-vanishing and proportional to the background torsion. In section V we show that the vacuum expectation value of the energy-momentum tensor on the physical vacuum in non trivial. In the section VI we derive an analytical form of the axial current and energy-momentum tensor on the physical vacuum, in the strong coupling limit. The last section is devoted to the discussion of the possible implications of the
fermion condensate and to the conclusions. In the appendix A, we report the computations of
the auxialiry tensor appearing in the calculus of the vacuum expectation value of the axial
current.

\section{Square-torsion theory}
Let us consider a spacetime $M$, endowed with a metric tensor $g_{\mu\nu}$. In this general framework, we consider the square-torsion modification of Einstein--Cartan gravity proposed in \cite{Fabbri:2012yg, VFS}, coupled with a Dirac field. The total Lagrangian $\mathcal{L}=\mathcal{L}_G+\mathcal{L}_M$ is the sum of the gravitational Lagrangian $\mathcal{L}_G$ and the matter Lagrangian $\mathcal{L}_M$. Where $\mathcal{L}_G$ is expressed as the sum
\begin{equation}\label{Lagrangian}
\mathcal{L}_G = \tilde{R}+\zeta T\,,
\end{equation}
with $\tilde{R}$ scalar curvature of the dynamical connection and $T:=T_{\mu\nu}^{\;\;\;\lambda}T^{\mu\nu}_{\;\;\;\lambda}$, with $T_{\mu\nu}^{\;\;\;\lambda}$ denoting the torsion of the dynamical connection. $\zeta$ is a suitable coupling constant. If the parameter $\zeta$ goes to zero, we obtain the standard Einstein-Sciama-Kibble theory again. Moreover the matter Lagrangian $\mathcal{L}_M$ is that of the Dirac field with spinorial covariant derivatives containing torsion: 
	\begin{equation}
		\mathcal{L}_M = \left[ \frac{i}{2} \left( \bar{\phi} \gamma^i D_i \phi - D_i \bar{\phi} \gamma^i \phi \right) - m \bar{\phi} \phi \right]\,.
	\end{equation}		
Denoting by $\gamma^A$ ($A=0,1,2,3$) the set of flat Dirac matrices, we set the curved Dirac matrices $\tilde\gamma^\mu = e^\mu_A\gamma^A$, where $e^A_\mu$ ($e^A_{\mu}e^\mu_B=\delta^A_B$ and $e^\mu_{A}e^A_\nu=\delta^\mu_\nu$) indicate the elements of a tetrad field associated with the metric $g_{\mu\nu}$. Indexes $A$ are raised and lowered by the Minkowski metric $\eta_{AB}$. Moreover, setting $S_{AB}:= \frac{1}{8}[\gamma_A,\gamma_B]$, the covariant derivative of the Dirac field $\phi$ is $D_{\mu} \phi =\frac{\partial \phi} {\partial x^\mu } + \omega_\mu ^{AB} S_ {AB} \phi$ and  $D_\mu \bar \phi= \frac{\partial \bar \phi} {\partial x^\mu } -\bar \phi \omega_\mu ^{AB} S_{AB}$, where $\bar \phi  = \phi^{\dagger} \gamma^0$ is the conjugate spinor and $\omega_ \mu ^{AB}$ is the spin connection. The consequent field equations are of the form \cite{VFS} 

\begin{subequations}\label{field_equations}
\begin{equation}\label{field_equations_1}
G_{\mu\nu} = \Sigma_{\mu\nu} + \frac{3\mathcal{Y}}{64}(\bar{\phi}\gamma_5\tilde{\gamma}^\tau\phi)(\bar{\phi}\gamma_5\tilde{\gamma}_\tau\phi)g_{\mu\nu},
\end{equation}
\begin{equation}\label{field_equations_2}
T_{\mu\nu}^{\;\;\;\lambda}
=\mathcal{Y}S_{\mu\nu}^{\;\;\;\lambda}
\end{equation}
and
\begin{equation}\label{field_equations_3}
i\tilde\gamma^{\lambda}D_{\lambda}\phi
-\frac{3\mathcal{Y}}{16}\left[(\bar{\phi}\phi)
+i(i\bar{\phi}\gamma_5\phi)\gamma_5\right]\phi-m\phi=0\,.
\end{equation}
\end{subequations}
Here,
\begin{equation}\label{eq: energia_impulso}
\Sigma_{\mu\nu} = \frac{i}{4}\/\left[ \bar\phi\tilde\gamma_{(\mu}D_{\nu)}\phi - \left(D_{(\nu}\bar\phi\right)\tilde\gamma_{\mu)}\phi \right]
\end{equation}
and
\begin{equation}\label{spin}
S_{\mu\nu}^{\;\;\;\lambda}=\frac{i}{2}\bar\phi\left\{\tilde\gamma^{\lambda},S_{\mu\nu}\right\}\phi
\end{equation}
are the symmetrized energy--momentum tensor and the spin tensor of the Dirac field $\phi$ respectively, $G_{\mu\nu}$ is the Einstein tensor induced by the metric $g_{\mu\nu}$ and $\mathcal{Y}$ denotes the constant $\mathcal{Y}=\frac{1}{1-4\zeta}$ ($\zeta\not= \frac{1}{4}$). The classical Dirac field $\phi$ sources the right hand side of the field equations \eqref{field_equations_1} and \eqref{field_equations_2}.  In the above equations $D_\mu$ denotes the spinorial covariant derivative induced by the Levi--Civita connection.

Now, let us consider a spatially flat Friedmann-Lemaître-Robertson-Walker (FLRW) metric, expressed as
\begin{equation}\label{metric}
ds^2 = dt^2 - C^2(t)\,dx^2 - C^2(t)\,dy^2 - C^2(t)\,dz^2
\end{equation}

Evaluating the field equations \eqref{field_equations} on the metric tensor \eqref{metric} and making use of the standard representation for the Dirac matrices $\gamma^A$, as done in \cite{VFS}, one has the following Dirac and Einstein equations:
\begin{subequations}\label{equazioni_Dirac}
\begin{equation}\label{eq:7a}
\dot\phi + \frac{3\dot C}{2C}\phi + im\gamma^0\phi -
\frac{3\mathcal{Y}i}{16}\/\left[ (\bar\phi\phi)\gamma^0 +i\/(i\bar\phi\gamma^5\phi)\gamma^0\gamma^5 \right]\phi =0
\end{equation}
\begin{equation}\label{eq:7b}
\dot{\bar\phi} + \frac{3\dot C}{2C}\bar\phi - im\bar{\phi}\gamma^0 + \frac{3\mathcal{Y}i}{16}\bar\phi\/\left[ (\bar\phi\phi)\gamma^0 +i\/(i\bar\phi\gamma^5\phi)\gamma^5\gamma^0 \right] =0
\end{equation}
\end{subequations}
and
\begin{subequations}\label{equazioni_Einstein}
\begin{equation}
3\frac{{\dot C}^2}{C^2} =
\frac{1}{2}m\bar\phi\phi - \frac{3\mathcal{Y}}{64}(\bar{\phi}\gamma^5\gamma^A\phi)(\bar{\phi}\gamma^5\gamma_A\phi)
\end{equation}
\begin{equation}
2\frac{\ddot C}{C} + \frac{{\dot C}^2}{C^2} =  \frac{3\mathcal{Y}}{64}(\bar{\phi}\gamma^5\gamma^A\phi)(\bar{\phi}\gamma^5\gamma_A\phi).
\end{equation}
\end{subequations}
By multiplyng the Eq. \eqref{eq:7a} on the left by $\overline{\phi}$ and the Eq. \eqref{eq:7b} on the right by $\phi$ and summing, we obtain:
\begin{equation}\label{eq:phibarra}
	\overline{\phi}\dot{\phi}+\dot{\overline{\phi}}\phi+\frac{3\dot{C}}{C}\overline{\phi}\phi+3\frac{\mathcal{Y}}{8}\left(i\overline{\phi}\gamma^5\phi\right)\overline{\phi}\gamma^0\gamma^5\phi=0\,.
\end{equation} Under the simplifying condition
\begin{equation}\label{condizione}
\bar\phi\gamma^5\phi =0
\end{equation}
from eqs. \eqref{eq:phibarra} we obtain the relation
\begin{equation}\label{eq: psibarrapsi}
\bar\phi\phi= \frac{Z}{C^3}
\end{equation}
with $Z=const$. In such a circumstance, by means of relations \eqref{condizione} and \eqref{eq: psibarrapsi}, it is easily seen that Dirac equations \eqref{equazioni_Dirac} admit general solution of the form
\begin{eqnarray}
\label{generalspinor}
&\phi=C^{-3/2}\left(\begin{tabular}{c}
$H_1e^{-i m t - i \xi(t)}$\\
$H_2e^{-im t  - i \xi(t)}$\\
$H_3e^{+im t + i \xi(t)}$\\
$H_4e^{+im t + i \xi(t)}$
\end{tabular}\right)
\end{eqnarray}
where $\xi(t)=\frac{3\mathcal{Y}K}{16}\int{\frac{dt}{C^3}}$ and $H_i$ are complex numbers which satisfy the relations
\begin{equation}
H_1^*H_1 + H_2^*H_2 - H_3^*H_3 - H_4^*H_4 = Z
\label{condizione1}
\end{equation}
and
\begin{equation}
H_1^*H_3 + H_2^*H_4 - H_3^*H_1 - H_4^*H_2 = 0
\label{condizione2}
\end{equation}
coming from eqs. \eqref{condizione} and \eqref{eq: psibarrapsi}. Moreover, the components of the spin speudo--vector $\bar\phi\gamma^5\gamma^A\phi$ are seen to be
\begin{equation} \label{eq: componenti}
\begin{split}
\bar\phi\gamma^5\gamma^0\phi&=-\frac{2\left(H_3H_1^*+H_4H_2^*\right)}{C^3}\cos \left( 2\xi(t)+2mt\right)\\
\bar\phi\gamma^5\gamma^1\phi&=-\frac{H_2H_1^*+H_1H_2^*+H_4H_3^*+H_3H_4^*}{C^3}\\
\bar\phi\gamma^5\gamma^2\phi&=\frac{i \left(H_2H_1^*-H_1H_2^*+H_4H_3^*-H_3H_4^*\right)}{C^3}\\
\bar\phi\gamma^5\gamma^3\phi&= -\frac{-2H_2H_2^*+2 H_3H_3^*+Z}{C^3}.
\end{split}
\end{equation}

We denote by $\breve{T}^{\rho}$ the components of the pseudo-vector $\bar\phi\gamma^5\tilde{\gamma}^\rho\phi$ as shown below
\begin{equation}
\begin{split}
\breve{T}^{0}=\bar\phi\gamma^5\tilde{\gamma}^0\phi\\
\breve{T}^{1}=\bar\phi\gamma^5\tilde{\gamma}^1\phi\\
\breve{T}^{2}=\bar\phi\gamma^5\tilde{\gamma}^2\phi\\
\breve{T}^{3}=\bar\phi\gamma^5\tilde{\gamma}^3\phi.\\
\end{split}
\end{equation}

\section{Dirac equation with torsion}

We wish now to study the dynamics of a quantized Dirac field $\psi$ in the torsionful curved background described above.
A standard choice of tetrads for the metric of Eq. \eqref{metric} is
\begin{equation}
e_{\mu}^{0}=\delta_{\mu}^{0}\,, \ \ \ \ \ e_{\mu}^{J} = C(t) \delta_{\mu}^{J}
\end{equation}
where the Kronecker symbol $\delta_{\mu}^{A}$ imposes a nonzero value of the tetrad fields $e_{\mu}^{A}$ for $A = \mu$ and $J = 1,2,3$. Moreover,   we  need the spin
connection $\omega_{\mu}^{AB}=e_{\nu}^{A}\Gamma_{\sigma\mu}^{\nu}e^{\sigma B}+e_{\nu}^{A}\partial_{\mu}e^{\nu B}$, which allows us to write
the spinorial covariant derivative,  $D_{\mu}\psi=\partial_{\mu}\psi+\Gamma_{\mu}\psi$,
where $\Gamma_{\mu}=\frac{1}{8}\omega_{\mu}^{AB}\left[\gamma_{A},\gamma_{B}\right]$.
The Dirac equation, with torsion in FLRW metric, is then (see also \cite{Torsioncostant}):
\begin{equation} \label{eq: Dirac_equation}
i\tilde{\gamma}^{\mu}D_{\mu}\psi-M\psi=-\frac{3\mathcal{Y}}{16}\breve{T}^{\rho}\tilde{\gamma}_{\rho}\gamma^{5}\psi\,.
\end{equation}
Here, $\breve{T}^{\rho}=\left(\bar{\phi}\gamma^{5}\tilde{\gamma}^{\rho}\phi\right)$,
where $\phi$ is the classical background Dirac field in the spatially flat FLRW metric (see the section above), and $M$ is the mass of the field $\psi$ (not to be confused with $m$, the mass of the background field $\phi$) .
The Dirac Equation reads\footnote{See for instance also \cite{FV1}}: 
\begin{equation}\label{d1}
\left(i\gamma^{0}\partial_{t}+\frac{3}{2}i\frac{\partial_{t}C(t)}{C(t)}\gamma^{0}+i\gamma^{1}\frac{\partial_{x}}{C(t)}+i\gamma^{2}\frac{\partial_{y}}{C(t)}+i\gamma^{3}\frac{\partial_{z}}{C(t)}-
M\right)\psi=-\frac{3\mathcal{Y}}{16}\breve{T}^{\rho}\tilde{\gamma}_{\rho}\gamma^{5}\psi\, \ .
\end{equation}
To solve the Dirac equation (\ref{d1}) we expand the Dirac field  $\psi(x)$ as
\begin{equation}
\psi(x)=\sum_{\lambda}\int d^{3}k\left(b_{\vec{k},\lambda}u_{\vec{k},\lambda}+d_{\vec{k}, \lambda}^{\dagger}v_{\vec{k},\lambda}\right)\,,\label{eq: spinore psi}
\end{equation}
and use the ansatz: $u_{\vec{k},\lambda}(t,\boldsymbol{x})=e^{i\boldsymbol{k}\cdot\boldsymbol{x}}\left(\begin{array}{c}
f_{k,\lambda}(t)\xi_{\lambda}(\hat{k})\\
g_{k,\lambda}(t)\lambda\xi_{\lambda}(\hat{k})
\end{array}\right)$ for positive energies, and $v_{\vec{k},\lambda}(t,\boldsymbol{x})=e^{i\boldsymbol{k}\cdot\boldsymbol{x}}\left(\begin{array}{c}
g_{k,\lambda}^{*}(t)\xi_{\lambda}(\hat{k})\\
-f_{k,\lambda}^{*}(t)\lambda\xi_{\lambda}(\hat{k})
\end{array}\right)$ for negative energies. Here we have introduced the helicity eigenspinors $\xi_{\lambda}(\hat{k})$, with helicity $\lambda =\pm 1$ and $\hat{k}$ is the unit vector defined as follow $\hat{k}\equiv\frac{\vec{k}}{\abs{\vec{k}}}$ . The operator coefficients $b_{\vec{k},\lambda}, d_{\vec{k},\lambda}$ satisfy the canonical anticommutation relations as usual. We now insert the ansatz for the positive energy solution $u_{\vec{k},\lambda}(t,\boldsymbol{x})$ in the Eq\eqref{d1} and multiply on the left by $ \xi^\dagger_\lambda(\hat{k}) $. For the right-hand side we obtain: 
\begin{equation*}
	-\frac{3\mathcal{Y}}{16}\breve{T}^\rho\xi^\dagger_\lambda(\hat{p})\tilde{\gamma}_\rho\left(\begin{array}{c}
		g_{k,\lambda}(t)\lambda\xi_{\lambda}(\hat{k})\\
		f_{k,\lambda}(t)\xi_{\lambda}(\hat{k})
	\end{array}\right)=-\frac{3\mathcal{Y}}{16}\left(\begin{array}{c}
		\lambda\breve{T}^0g_{k,\lambda}(t)+C(t)\breve{T}^if_{k,\lambda}(t)\xi^\dagger_\lambda(\hat{k})\sigma^i\xi_{\lambda}(\hat{k})\\
		-C(t)\breve{T}^ig_{k,\lambda}(t)\xi^\dagger_\lambda(\hat{k})\sigma^i\xi_{\lambda}(\hat{k}) -\breve{T}^0 f_{k,\lambda}(t)
	\end{array}\right)\,,
\end{equation*}
using the relation $\xi^\dagger \, \sigma^i \, \xi_\lambda = \lambda \, \frac{\vec{k}}{\abs{\vec{k}}} = \lambda\hat{k}$ and remembering that $\lambda \lambda = \mathbb{I}_{2}$, the Eq. (\ref{d1})  becomes \footnote{We use the defining relation of the helicity eigenspinors
$\vec{\sigma}\cdot\vec{k}\xi_{\lambda}=\lambda k\xi_{\lambda}$ }:
\begin{equation}\label{d2}
i\partial_{t}\left(\begin{array}{c}
f_{\vec{k},\lambda}(t)\\
g_{\vec{k},\lambda}(t)
\end{array}\right)=\left(\begin{array}{cc}
-\frac{3}{2}i\frac{\partial_{t}C}{C}+M-\frac{3\mathcal{Y}}{16} C(t)\lambda\breve{T}^{i}\hat{k}^{i} & \frac{k}{C(t)}-\frac{3\mathcal{Y}}{16}\lambda\breve{T}^{0}\\
\frac{k}{C(t)}-\frac{3\mathcal{Y}}{16}\lambda\breve{T}^{0} & -\frac{3}{2}i\frac{\partial_{t}C}{C}-M-\frac{3\mathcal{Y}}{16} C(t)\lambda\breve{T}^{i}\hat{k}^{i}
\end{array}\right)\left(\begin{array}{c}
f_{\vec{k},\lambda}(t)\\
g_{\vec{k},\lambda}(t)
\end{array}\right).
\end{equation}
In order to eliminate the derivative term of the scale factor, we rescale the functions as follows:
\begin{equation} \label{eq: Phi_gamma}
\Phi_{\vec{k},\lambda}\equiv C^{\frac{3}{2}}f_{\vec{k}, \lambda \,,}\qquad\gamma_{\vec{k},\lambda}\equiv C^{\frac{3}{2}}g_{\vec{k},\lambda}\,,
\end{equation}
so that eq.(\ref{d2}) becomes:
\begin{equation}
i\partial_{t}\left(\begin{array}{c}
\Phi_{\vec{k},\lambda}(t)\\
\gamma_{\vec{k},\lambda}(t)
\end{array}\right)=\left(\begin{array}{cc}
M-\frac{3\mathcal{Y}}{16} C(t)\lambda\breve{T}^{i}\hat{k}^{i} & \frac{k}{C(t)}-\frac{3\mathcal{Y}}{16}\lambda\breve{T}^{0}\\
\frac{k}{C(t)}-\frac{3\mathcal{Y}}{16}\lambda\breve{T}^{0} & -M-\frac{3\mathcal{Y}}{16} C(t)\lambda\breve{T}^{i}\hat{k}^{i}
\end{array}\right)\left(\begin{array}{c}
\Phi_{\vec{k},\lambda}(t)\\
\gamma_{\vec{k},\lambda}(t)
\end{array}\right)\,.
\end{equation}
Eqs.(\ref{condizione1}) and  (\ref{condizione2}) leave two parameters free. Since the component $\breve{T}^{0}$ is defined as follows:
\begin{equation}
\breve{T}^{0}\equiv-\frac{2\left(H_{3}H_{1}^{*}+H_{4}H_{2}^{*}\right)}{C^{3}(t)}\cos\left(2\xi(t)+2mt\right)\,,
\end{equation}
it is possible to choose $H_{3}H_{1}^{*}+H_{4}H_{2}^{*}=0$ to simplify
the calculations. This choice satisfies eqs.(\ref{condizione1}) and (\ref{condizione2}). 
The components $\breve{T}^{i}$ are of the form $\breve{T}^{i}=\frac{constant}{C^{3}(t)}$, therefore
the   time dependence of $\breve{T}^{i}$ is due to the term
  $C(t).$ To progress further it is convenient to introduce the
conformal time $d\tau = \frac{dt}{C(t)}$. Denoting with $G(\tau)\equiv\breve{T}^{i} (\tau) \hat{k}^{i}$,
we have:
\begin{equation}
i\partial_{\tau}\begin{pmatrix}
\Phi_{\vec{k},\lambda}(\tau)\\
\gamma_{\vec{k},\lambda}(\tau)
\end{pmatrix}=
\begin{pmatrix}
MC-\frac{3\mathcal{Y}}{16}\lambda C^{2}G(\tau) & k\\
k & -MC-\frac{3\mathcal{Y}}{16}\lambda C^{2}G(\tau)
\end{pmatrix}
\begin{pmatrix}
\Phi_{\vec{k},\lambda}(\tau)\\
\gamma_{\vec{k},\lambda}(\tau)
\end{pmatrix}=
\mathscr{B}(\tau)
\begin{pmatrix}
\Phi_{\vec{k}}(\tau)\\
\gamma_{\vec{k}}(\tau)
\end{pmatrix}\,.
\end{equation}
The system can now be easily solved, leading to
\begin{equation} \label{eq: solution_Phi_gamma}
\begin{array}{c}
\Phi_{\vec{k},\lambda}(\tau)=E_{\vec{k}}e^{-i\omega_{k}\tau}e^{i\int_{0}^{\tau}\frac{3\mathcal{Y}}{16}\lambda C^{2}(t)G(t)dt}\,,\\
\gamma_{\vec{k},\lambda}(\tau)=\frac{E_{\vec{k}}}{\omega_{k}+MC}ke^{-i\omega_{k}\tau}e^{i\int_{0}^{\tau}\frac{3\mathcal{Y}}{16}\lambda C^{2}(t)G(t)dt}\,,
\end{array}
\end{equation}
where $E_{\vec{k}}=\frac{\omega_{k}+MC}{(2\pi)^{\frac{3}{2}}\sqrt{k^{2}+(\omega_{k}+MC)^{2}}}$
is obtained by the normalisation condition $\left|\Phi_{\vec{k}}\right|^{2}+\left|\gamma_{\vec{k}}\right|^{2}=\frac{1}{(2\pi)^{\frac{3}{2}}}\,,$
moreover $\omega_{k}$ is given by $\omega_{k}=\sqrt{k^{2}+M^{2}C^{2}}\,.$\\
\\
For simplicity, we consider  the case with $M\rightarrow0$. In this case, $f_{\vec{k},\lambda}$ and $g_{\vec{k},\lambda}$
assume  the following form:
\begin{equation}
\left\{ \begin{array}{c}
f_{\vec{k},\lambda} (\tau)=\frac{C^{-\frac{3}{2}}}{(2\pi)^{\frac{2}{3}}}\frac{1}{\sqrt{2}}e^{-i\omega_{k}\tau}e^{i\int_{0}^{\tau}\frac{3\mathcal{Y}}{16}\lambda C^{2}(t)G(t)dt}\,,\\
g_{\vec{k},\lambda}(\tau)=\frac{C^{-\frac{3}{2}}}{(2\pi)^{\frac{2}{3}}}\frac{1}{\sqrt{2}}e^{-i\omega_{k}\tau}e^{i\int_{0}^{\tau}\frac{3\mathcal{Y}}{16}\lambda C^{2}(t)G(t)dt}\,.
\end{array}\right.
\end{equation}
\\
Let us now denote with $H$ the field Hamiltonian of $\psi$, as defined by the time component of the related energy momentum tensor $T_{tt}$. Due to the torsion term the Hamiltonian is not diagonal in the free field creation operators and takes the form $H\equiv\int d^{3}k\,\bar{b}_{\vec{k}}^{\dagger}\mathscr{H}_{\vec{k}}(t)\bar{b}_{\vec{k}}$,
with $\bar{b}_{\vec{k}}^{\dagger}\equiv\left(b_{\vec{k},+}^{\dagger},b_{\vec{k},-}^{\dagger},d_{\vec{k},+},d_{\vec{k},-}\right)$,
and $\mathscr{H}_{\vec{k}}(t)$   defined by
\begin{equation}
\mathscr{H}_{\vec{k}}(t)\equiv\left(\begin{array}{cccc}
P_{\vec{k},+}(t) & 0 & 0 & V_{\vec{k}}^{*}(t)\\
0 & P_{\vec{k},-} & U_{\vec{k}}^{*}(t) & 0\\
0 & U_{\vec{k}}(t) & Q_{\vec{k},+}(t) & 0\\
V_{\vec{k}}(t) & 0 & 0 & Q_{\vec{k},-}(t)
\end{array}\right).
\end{equation}
\\
where the coefficients $P_{\vec{k},\lambda}(t),  Q_{\vec{k},\lambda}(t), U_{\vec{k}}(t), V_{\vec{k}}(t)$ are given, in the limit $M\rightarrow 0$, by
\begin{align}
    P_{\vec{k},\lambda}(t) &=  - \frac{3\mathcal{Y}}{8} \lambda \, \breve{T}^{i} \hat{k}^i + kC(t)-i\frac{3}{2}\frac{\dot{C}(t)}{C(t)}\,, \\
    Q_{\vec{k},\lambda}(t) &= - \frac{3\mathcal{Y}}{8} \lambda \, \breve{T}^{i} \hat{k}^i - kC(t)-i\frac{3}{2}\frac{\dot{C}(t)}{C(t)}\,, \\
    U_{\vec{k}}(t) &= - \frac{3\mathcal{Y}}{8} \mu_{\vec{k}} (t) \, C(t) e^{-2i \omega_k t}\,,\\
    V_{\vec{k}}(t) &= - \frac{3\mathcal{Y}}{8} \mu_{\vec{k}}^* (t) \, C(t) e^{-2i \omega_k t}\,.
\end{align}
The quantity $\mu_{\vec{k}}(t)$ is defined by the following relationship:
\begin{align} \label{eq: mu}
    \mu_{\vec{k}} (t) &\equiv \varepsilon^i (\hat{k}) \breve{T}^i= \notag \\
    &= +\breve{T}^1 \left[ \frac{1+\cos \theta_k}{2} - e^{-2i\phi_k} \left( \frac{1-\cos \theta_k}{2} \right) \right] +\notag \\
    &\quad - i \breve{T}^2 \left[ \frac{1+\cos \theta_k}{2} + e^{-2i\phi_k} \left( \frac{1-\cos \theta_k}{2} \right) \right]+ \notag \\
    &\quad - \breve{T}^3 e^{-i\phi_k} \sin \theta_k \,,
\end{align}
where $\theta_k$ and $\phi_k$ denote the angles formed by the unit vector $\hat{k}\equiv \left( \sin \theta \cos \phi, \sin \theta \sin \phi, \cos \theta \right)$ and we have further defined the vector $\vec{\varepsilon}(\hat{k})\equiv \xi_{+} ^\dagger (\hat{k}) \vec{\sigma}\xi_{-}(\hat{k})$. \\
 $H$   can be diagonalised by means of the Bogoliubov
transformation
\begin{equation} \label{eq: transformations}
\left\{ \begin{array}{c}
B_{\vec{k},+}=M_{\vec{k},++}b_{\vec{k},+}+N_{\vec{k},+-}d_{\vec{k},-\,,}^{\dagger}\\
B_{\vec{k},-}=M_{\vec{k},--}b_{\vec{k},-}+N_{\vec{k},-+}d_{\vec{k},+\,,}^{\dagger}\\
D_{\vec{k},+}^{\dagger}=N_{\vec{k},-+}^{*}b_{\vec{k},-}-M_{\vec{k},--}d_{\vec{k},+\,,}^{\dagger}\\
D_{\vec{k},-}^{\dagger}=N_{\vec{k},+-}^{*}b_{\vec{k},+}-M_{\vec{k},++}d_{\vec{k},-\,,}^{\dagger}
\end{array}\right.
\end{equation}
for appropriate choices of the coefficients $M_{\vec{k},\lambda \lambda'}, N_{\vec{k},\lambda \lambda'}$.
The transformation can be recast in terms of the generator $\mathscr{R}(t)$
\begin{equation}
\begin{array}{c}
B_{\vec{k},\pm}(t)=\mathscr{R}^{-1}(t)b_{\vec{k},\pm}\mathscr{R}(t)\,,\\
D_{\vec{k},\pm}(t)=\mathscr{R}^{-1}(t)d_{\vec{k},\pm}\mathscr{R}(t)\,.
\end{array}
\end{equation}
Below are shown the Bogoliubov coefficients $M_{\vec{k}}$ and $N_{\vec{k}}$, in the general case, which appear in equation \eqref{eq: transformations}
\begin{align}
   M_{\vec{k},++} &\equiv \frac{\left( P_{\vec{k},+}(t)- Q_{\vec{k},-}(t) \right)+\sqrt{\left(P_{\vec{k},+}(t)- Q_{\vec{k},-}(t) \right)^2 +4 \left|V_{\vec{k}}\right|^2 }}{\sqrt{\left(P_{\vec{k},+}(t)- Q_{\vec{k},-}(t)+\sqrt{\left( P_{\vec{k},+}(t)- Q_{\vec{k},-}(t) \right)^2+4 \left|V_{\vec{k}}\right|^2}  \right)^2+4 \left|V_{\vec{k}}\right|^2}}  \,, \\
M_{\vec{k},--} &\equiv \frac{\left( P_{\vec{k},-}(t)- Q_{\vec{k},+}(t) \right)+\sqrt{\left(P_{\vec{k},-}(t)- Q_{\vec{k},+}(t) \right)^2 +4 \left|U_{\vec{k}}\right|^2 }}{\sqrt{\left(P_{\vec{k},-}(t)- Q_{\vec{k},+}(t)+\sqrt{\left( P_{\vec{k},-}(t)- Q_{\vec{k},+}(t) \right)^2+4 \left|U_{\vec{k}}\right|^2}  \right)^2+4 \left|U_{\vec{k}}\right|^2}}  \,, \\
  N_{\vec{k},+-} &\equiv \frac{2  V^*_{\vec{k}}(t)  }{\sqrt{\left(P_{\vec{k},+}(t)- Q_{\vec{k},-}(t)+\sqrt{\left( P_{\vec{k},+}(t)- Q_{\vec{k},-}(t) \right)^2+4 \left|V_{\vec{k}}\right|^2}  \right)^2+4 \left|V_{\vec{k}}\right|^2}} \,,  \\
 N_{\vec{k},-+} &\equiv \frac{2  U^*_{\vec{k}}(t)  }{\sqrt{\left(P_{\vec{k},-}(t)- Q_{\vec{k},+}(t)+\sqrt{\left( P_{\vec{k},-}(t)- Q_{\vec{k},+}(t) \right)^2+4 \left|U_{\vec{k}}\right|^2}  \right)^2+4 \left|U_{\vec{k}}\right|^2}}  \,.
\end{align}
\section{The expectation value of the spin operator on the vacuum}

The vacuum annihilated by the operators $B_{\vec{k},\pm}(t)$ and $D_{\vec{k},\pm}(t)$
denoted by $\left|0_{c}(t)\right\rangle $ is linked to the vacuum
$\left|0\right\rangle $ annihilated by the operators  $b_{\vec{k},\pm}$
and $d_{\vec{k},\pm}$ by the relation:
\\
\begin{equation}
\left|0_{c}(t)\right\rangle =\mathscr{R}^{-1}\left|0\right\rangle \,.
\end{equation}

We want to determine the following expectation value $L^{\mu}\equiv$$\left\langle 0_{c}(t_0)\left|\bar{\psi}\gamma^{5}\tilde{\gamma}^{\mu}\psi\right|0_{c}(t_0)\right\rangle $, where $t_0$ is fixed.
\\
Inserting the expression of the fields  in eq.(\ref{eq: spinore psi}),
we found that the component $L^{0} = 0$. In fact, one has:
\begin{align*}
L^{0} & =-\sum_{\lambda,\lambda'}\int d^{3}k\int d^{3}q\frac{e^{i\left(\vec{q}-\vec{k}\right)\cdot\vec{x}}}{C(t)}\left\{ \Delta_{\vec{k},\vec{q},\lambda,\lambda'}\left\langle 0_{c}(t_0)\left|b_{\vec{k},\lambda}^{\dagger}b_{\vec{q},\lambda'}\right|0_{c}(t_0)\right\rangle + \Lambda_{\vec{k},\vec{q},\lambda,\lambda'}\left\langle 0_{c}(t_0)\left|b_{\vec{k},\lambda}^{\dagger}d_{\vec{q},\lambda'}^{\dagger}\right|0_{c}(t_0)\right\rangle +\right.\\
 & \left.+\Xi_{\vec{k},\vec{q},\lambda,\lambda'}\left\langle 0_{c}(t_0)\left|d_{\vec{k},\lambda}b_{\vec{q},\lambda'}\right|0_{c}(t_0)\right\rangle +\Pi_{\vec{k},\vec{q},\lambda,\lambda'}\left\langle 0_{c}(t_0)\left|d_{\vec{k},\lambda}d_{\vec{q},\lambda'}^{\dagger}\right|0_{c}(t_0)\right\rangle \right\} \,,
\end{align*}

where the coefficients are explicitly:
\begin{align}
    \Delta_{\vec{k},\vec{q},\lambda,\lambda'} &= \xi_{\lambda}^\dagger (\hat{k}) \xi_{\lambda'}(\hat{q}) \left( \lambda' f_{\vec{k},\lambda}^* g_{\vec{q},\lambda'} + \lambda g_{\vec{k},\lambda}^* f_{\vec{q},\lambda'} \right), \\
    \Lambda_{\vec{k},\vec{q},\lambda,\lambda'} &= \xi_{\lambda}^\dagger (\hat{k}) \xi_{\lambda'}(\hat{q}) \left( -\lambda' f_{\vec{k},\lambda}^* f_{\vec{q},\lambda'}^* + \lambda g_{\vec{k},\lambda}^* g_{\vec{q},\lambda'}^* \right), \\
    \Xi_{\vec{k},\vec{q},\lambda,\lambda'} &= \xi_{\lambda}^\dagger (\hat{k}) \xi_{\lambda'}(\hat{q}) \left( \lambda' g_{\vec{k},\lambda} g_{\vec{q},\lambda'} - \lambda f_{\vec{k},\lambda} f_{\vec{q},\lambda'} \right), \\
    \Pi_{\vec{k},\vec{q},\lambda,\lambda'} &= \xi_{\lambda}^\dagger (\hat{k}) \xi_{\lambda'}(\hat{q}) \left( -\lambda' g_{\vec{k},\lambda} f_{\vec{q},\lambda'}^* - \lambda f_{\vec{k},\lambda} g^*_{\vec{q},\lambda'} \right).
\end{align}

Using the notation $\bar{\lambda}=-\lambda$ we have the following relationship:


\begin{equation}
 L^{0}=-\sum_{\lambda}\int d^{3}k\left\{ \lambda\frac{C^{-4}(t)}{(2\pi)^{3}}\left(\left|N_{\vec{k},\lambda\bar{\lambda}}(t_0)\right|^{2}-\left|M_{\vec{k},\bar{\lambda}\bar{\lambda}}(t_0)\right|^{2}\right)\right\} \,.
\end{equation}
\\
Since the coefficients $M_{\vec{k}}$ and $N_{\vec{k}}$ are linked by the relation
\begin{equation}
\left|M_{\vec{k},\bar{\lambda}\bar{\lambda}}\right|^{2}=1-\left|N_{\vec{k},\bar{\lambda}\lambda}\right|^{2}\,,
\end{equation}
the quantity $L^{0}  = -\sum_{\lambda}\int d^{3}k\left\{ \lambda\frac{C^{-4}(t)}{(2\pi)^{3}}\left(\left|N_{\vec{k},\lambda\bar{\lambda}}(t_0)\right|^{2}+\left|N_{\vec{k},\bar{\lambda}\lambda}(t_0)\right|^{2} - 1\right)\right\}$ vanishes by symmetry, since the summand is odd in $\lambda$.

The quantity $\vec{L}$ is given by:


\begin{align*}
\vec{L} & =-\sum_{\lambda,\lambda'}\int d^{3}k\int d^{3}q\frac{e^{i\left(\vec{q}-\vec{k}\right)\cdot\vec{x}}}{C(t)}\left\{ \vec{\Delta}_{\vec{k},\vec{q},\lambda,\lambda'}\left\langle 0_{c}(t_0)\left|b_{\vec{k},\lambda}^{\dagger}b_{\vec{q},\lambda'}\right|0_{c}(t_0)\right\rangle +\vec{\Lambda}_{\vec{k},\vec{q},\lambda,\lambda'}\left\langle 0_{c}(t_0)\left|b_{\vec{k},\lambda}^{\dagger}d_{\vec{q},\lambda'}^{\dagger}\right|0_{c}(t_0)\right\rangle +\right.\\
 & \left.+\vec{\Xi}_{\vec{k},\vec{q},\lambda,\lambda'}\left\langle 0_{c}(t_0)\left|d_{\vec{k},\lambda}b_{\vec{q},\lambda'}\right|0_{c}(t_0)\right\rangle +\vec{\Pi}_{\vec{k},\vec{q},\lambda,\lambda'}\left\langle 0_{c}(t_0)\left|d_{\vec{k},\lambda}d_{\vec{q},\lambda'}^{\dagger}\right|0_{c}(t_0)\right\rangle \right\} \,,
\end{align*}

where the coefficients are:
\begin{align}
    \vec{\Delta}_{\vec{k},\vec{q},\lambda,\lambda'} &=\left( \xi_{\lambda}^\dagger (\hat{k}) \vec{\sigma} \xi_{\lambda'}(\hat{q}) \right) \left[ f^*_{\vec{k},\lambda}f_{\vec{q},\lambda'} + \lambda\lambda'\, g_{\vec{k},\lambda}^* g_{\vec{q},\lambda'} \right]\,, \\
   \vec{\Lambda}_{\vec{k},\vec{q},\lambda,\lambda'} &=\left( \xi_{\lambda}^\dagger (\hat{k}) \vec{\sigma} \xi_{\lambda'}(\hat{q}) \right) \left[ f_{\vec{k},\lambda}^*g_{\vec{q},\lambda'}^* - \lambda\lambda'\,g_{\vec{k},\lambda}^* f_{\vec{q},\lambda'}^* \right]\,, \\
   \vec{\Xi}_{\vec{k},\vec{q},\lambda,\lambda'} &= \left( \xi_{\lambda}^\dagger (\hat{k}) \vec{\sigma} \xi_{\lambda'}(\hat{q}) \right) \left[ f_{\vec{k},\lambda}g_{\vec{q},\lambda'} - \lambda\lambda'\,g_{\vec{k},\lambda} f_{\vec{q},\lambda'} \right]\,, \\
    \vec{\Pi}_{\vec{k},\vec{q},\lambda,\lambda'} &= \left( \xi_{\lambda}^\dagger (\hat{k}) \vec{\sigma} \xi_{\lambda'}(\hat{q}) \right) \left[ g_{\vec{k},\lambda}g^*_{\vec{q},\lambda'} + \lambda\lambda'\, f_{\vec{k},\lambda}f^*_{\vec{q},\lambda'} \right]\,. \\
\end{align}

%

which can be rewritten as follows:

\begin{equation}
	\begin{split}
		\vec{L} = -\frac{C^{-4}(t)}{(2\pi)^{3}}\sum_{\lambda}\int d^{3}k\Bigg\{ & \lambda\hat{k}\left(\left|N_{\vec{k},\lambda\bar{\lambda}}(t_0)\right|^{2}+\left|M_{\vec{k},\bar{\lambda}\bar{\lambda}}(t_0)\right|^{2}\right) +\\
		& -\left(\delta_{\lambda,+}\delta_{\bar{\lambda},-}\vec{\varepsilon}(\hat{k})+\delta_{\lambda,-}\delta_{\bar{\lambda},+}\vec{\varepsilon}^{*}(\hat{k})\right)\left[\left(e^{+2i\omega_{k}\tau}\right)N_{\vec{k},\lambda\bar{\lambda}}^{*}(t_0)M_{\vec{k},\lambda \lambda}(t_0)+\left(e^{-2i\omega_{k}\tau}\right)N_{\vec{k},\bar{\lambda}\lambda}(t_0)M_{\vec{k},\bar{\lambda}\bar{\lambda}}(t_0)\right]\Bigg\}\,.
		\label{eq:vecL}
	\end{split}
\end{equation}


The quantity $\sum_{\lambda}\left[\lambda\left(\left|N_{\vec{k},\lambda\bar{\lambda}}\right|^{2}+\left|M_{\vec{k},\bar{\lambda}\bar{\lambda}}\right|^{2}\right)\right]=2\left(\left|N_{\vec{k},+-}\right|^{2}-\left|N_{\vec{k},-+}\right|^{2}\right)$
was calculated using the Mathematica programme.
Defining
$G\equiv \breve{T}^{i}\hat{k}^{i}$, we have:
\begin{align*}
\left(\left|N_{\vec{k},+-}\right|^{2}\pm \left|N_{\vec{k},-+}\right|^{2}\right)  =9\mathcal{Y}^2 |\mu_{\vec{k}}|^{2} &\left( \frac{1}{18 \mathcal{Y}^2
|\mu_{\vec{k}}|^{2}-2\left(3 \mathcal{Y}G-8k\right)\left[ -3 \mathcal{Y}G +8  k +\sqrt{ \left( 3\mathcal{Y}G-8k\right)^2+9\mathcal{Y}^2|\mu_{\vec{k}}|^{2}}\right] }\right.+\\
 & \left.\pm \frac{1}{ 18  \mathcal{Y}^2
 	|\mu_{\vec{k}}|^{2}+2\left(3 \mathcal{Y}G+8k\right)\left[ +3 \mathcal{Y}  G +8  k +\sqrt{  \left( 3\mathcal{Y}G+8k\right)^2+9\mathcal{Y}^2|\mu_{\vec{k}}|^{2}}\right] }\right)\,,
\end{align*}

while the product of the mixed terms is given by:

\begin{equation}
	\begin{aligned}
		N_{\vec{k},+-}^*(t_0) M_{\vec{k},++}(t_0) &= -\frac{3\mathcal{Y}}{2}  \frac{\mu^*_{\vec{k}}(t_0)\,e^{+2i \omega_{\vec{k}}\tau}}{\sqrt{\left[\left(3\mathcal{Y}G-8k\right)^2+9\mathcal{Y}^2|\mu_{\vec{k}}|^{2} \right]}} \\
		N_{\vec{k},-+}^*(t_0) M_{\vec{k},--}(t_0) &= -\frac{3\mathcal{Y}}{2}  \frac{\mu_{\vec{k}}(t_0)\,e^{+2i \omega_{\vec{k}}\tau}}{\sqrt{\left[\left(3\mathcal{Y}G+8k\right)^2+9\mathcal{Y}^2|\mu_{\vec{k}}|^{2} \right]}} .
	\end{aligned}
\end{equation}

The equation \eqref{eq:vecL} is the general form of the vacuum expectation value of the axial current expressed in terms of the coefficients of Bogoliubov.

\section{Vacuum expectation value of the energy-momentum tensor}


Let us calculate the expectation value on the vacuum $\langle 0_{c}(t_0)|$ of the free part of energy-momentum tensor \begin{equation}
	\label{eq:T^QFT}
T^{\mu \nu}_{QFT}\equiv \frac{i}{4} \left[\overline{\psi} \tilde{\gamma}^{(\mu} D^{\nu)} \psi -D^{(\mu} \overline{\psi}\tilde{\gamma}^{\nu)} \psi \right] \,.
\end{equation}
 To do this, we first calculate the explicit form of this tensor. We substitute the Dirac field expansion $\psi(x)$ given in Eq.\eqref{eq: spinore psi} into the Eq.\eqref{eq: energia_impulso}. By introducing the auxiliary tensor through the following expression:

\begin{equation}
	\Omega_{\mu\nu}(A,B)=\overline{A}\tilde{\gamma}_\mu D_\nu B+\overline{A}\tilde{\gamma}_\nu D_\mu B - D_\mu \overline{A} \tilde{\gamma}_\nu B- D_\nu \overline{A} \tilde{\gamma}_\mu B \, ,
\end{equation}
it can be shown that the free part of energy-momentum tensor, as derived by the lagrangian density
\begin{equation}
\mathcal{L}_0=\dfrac{i \sqrt{-g}}{2}\overline{\psi}\left(\tilde{\gamma}^\mu D_\mu -M\right) \psi\,,
\end{equation}
 where $D_\mu$ denotes the spinorial covariant derivative, has the following form:
\begin{align}
	T^{QFT}_{\mu\nu} = \frac{i}{2} \sum_{\lambda,\lambda'} \int d^3k \, d^3q \left[ b_{\vec{k},\lambda}^\dagger b_{\vec{q},\lambda'} \Omega_{\mu\nu}(u_{\vec{k},\lambda},u_{\vec{q},\lambda'}) + b_{\vec{k},\lambda}^\dagger d_{-\vec{q},\lambda'}^\dagger \Omega_{\mu\nu}(u_{\vec{k},\lambda},v_{\vec{q},\lambda'}) + \right. \notag \\
	\left. + b_{-\vec{k},\lambda} d_{\vec{q},\lambda'} \Omega_{\mu\nu}(v_{\vec{k},\lambda},u_{\vec{q},\lambda'}) + d_{-\vec{k},\lambda} d_{-\vec{q},\lambda'}^\dagger \Omega_{\mu\nu}(v_{\vec{k},\lambda},v_{\vec{q},\lambda'}) \right]\,, \label{eq: T_munu}
\end{align}
where the functions $u_{\vec{k},r}$ and $v_{\vec{k},r}$ are the solutions of the Dirac equation \eqref{eq: Dirac_equation}, for positive and negative energy respectively, while $b_{\vec{k},r}$ and $d_{\vec{k},r}$ the usual annihilation operators that satisfy the canonical anticommutation relations. The application of normal ordering to Eq.\eqref{eq: T_munu} is to modify only the last term. In fact, taking into consideration the fermionic nature of this field, the relation is obtained:

\begin{align}
		:T^{QFT}_{\mu\nu}: = \frac{i}{2} \sum_{\lambda,\lambda'} \int d^3k \, d^3q \left[ b_{\vec{k},\lambda}^\dagger b_{\vec{q},\lambda'} \Omega_{\mu\nu}(u_{\vec{k},\lambda},u_{\vec{q},\lambda'}) + b_{\vec{k},\lambda}^\dagger d_{-\vec{q},\lambda'}^\dagger \Omega_{\mu\nu}(u_{\vec{k},\lambda},v_{\vec{q},\lambda'}) + \right. \notag \\
	\left. + b_{-\vec{k},\lambda} d_{\vec{q},\lambda'} \Omega_{\mu\nu}(v_{\vec{k},\lambda},u_{\vec{q},\lambda'}) - d_{-\vec{q},\lambda'}^\dagger d_{-\vec{k},\lambda}  \Omega_{\mu\nu}(v_{\vec{k},\lambda},v_{\vec{q},\lambda'}) \right]\,. \label{eq: normal_ordering_T_munu}
\end{align}

In order to calculate the expectation value on the vacuum of the energy-momentum tensor, the way the annihilators behave on this vacuum is shown below:
\begin{equation}
	\begin{aligned}
		\left\langle 0_{c}(t_0) \left| b_{\vec{k},\lambda}^{\dagger} b_{\vec{q},\lambda'} \right| 0_{c}(t_0) \right\rangle
		&= \delta^3 \left( \vec{k} - \vec{q} \right) \delta_{\lambda \lambda'} |N_{\vec{k},\lambda \overline{\lambda}}(t_0)|^2 \,, \\
		\left\langle 0_{c}(t_0) \left| b_{\vec{k},\lambda}^{\dagger} d_{\vec{q},\lambda'}^\dagger \right| 0_{c}(t_0) \right\rangle
		&= -\delta^3 \left( \vec{k} - \vec{q} \right) \delta_{\lambda \overline{\lambda'}} \left( N_{\vec{k},\lambda \overline{\lambda}}(t_0) \right)^* M_{\vec{k},\lambda \lambda}(t_0) \,, \\
		\left\langle 0_{c}(t_0) \left| d_{\vec{k},\lambda} b_{\vec{q},\lambda'} \right| 0_{c}(t_0) \right\rangle
		&= -\delta^3 \left( \vec{k} - \vec{q} \right) \delta_{\lambda \overline{\lambda'}} N_{\vec{k},\overline{\lambda} \lambda}(t_0) \left( M_{\vec{k},\overline{\lambda} \overline{\lambda}}(t_0) \right)^* \,, \\
		\left\langle 0_{c}(t_0) \left| d_{\vec{k},\lambda}^{\dagger} d_{\vec{q},\lambda'} \right| 0_{c}(t_0) \right\rangle
		&= \delta^3 \left( \vec{k} - \vec{q} \right) \delta_{\lambda \lambda'} |N_{\vec{k}, \overline{\lambda} \lambda }(t_0)|^2 \,.
	\end{aligned}
	\label{eq: centered_numbered_equations}
\end{equation}
Therefore, the following relationship is obtained:

\begin{align}
	\left\langle 0_{c}(t_0) \left| :T^{QFT}_{\mu\nu}: \right| 0_{c}(t_0) \right\rangle= \frac{i}{2} \sum_{\lambda} \int d^3k \Big[&|N_{\vec{k},\lambda \overline{\lambda}}(t_0)|^2 \Omega_{\mu\nu}(u_{\vec{k},\lambda}(t),u_{\vec{k},\lambda}(t))- \left( N_{\vec{k},\lambda \overline{\lambda}} (t_0) \right)^* M_{\vec{k},\lambda \lambda} (t_0) \Omega_{\mu\nu}(u_{\vec{k},\lambda}(t),v_{\vec{q},\overline{\lambda}}(t)) \notag \\
	&- N_{\vec{k},\overline{\lambda} \lambda} (t_0) \left( M_{\vec{k},\overline{\lambda} \overline{\lambda}}(t_0) \right)^* \Omega_{\mu\nu}(v_{\vec{k},\lambda}(t),u_{\vec{q},\overline{\lambda}}(t)) - |N_{\vec{k},\overline{\lambda} \lambda }(t_0)|^2 \Omega_{\mu\nu}(v_{\vec{k},\lambda}(t),v_{\vec{q},\lambda}(t)) \Big]\,. \label{eq: normal_ordering_T_munu}
\end{align}
The results for the calculation of the components of the auxiliary tensor ${\Omega_{tt}}$, such that $\mu=0$ and $\nu=0$, evaluated for appropriate combinations of the solutions of the Dirac equation $u_{\vec{p},\lambda}$ and $v_{\vec{p},\lambda}$, are given below:
\begin{align} \label{eq: Omega_tt_def}
	\Omega_{tt}(u_{\vec{k},\lambda},u_{\vec{k},\lambda}) &\equiv 2 \left[\overline{u}_{\vec{k},\lambda} \tilde{\gamma}_t D_t u_{\vec{k},\lambda} - \left( D_t \overline{u}_{\vec{k},\lambda} \right) \tilde{\gamma}_t u_{\vec{k},\lambda} \right]= \notag \\
	&= 2 C^{-3}(t) \left[ \Phi^*_{\vec{k},\lambda}(t) \overleftrightarrow{\partial_t} \Phi_{\vec{k},\lambda}(t) + \gamma^*_{\vec{k},\lambda}(t) \overleftrightarrow{\partial_t} \gamma_{\vec{k},\lambda}(t)\right]\,,
\end{align}
with $\Phi_{\vec{k},\lambda}$ and $\gamma
_{\vec{k},\lambda}$ given by Eq.\eqref{eq: solution_Phi_gamma}. More details about Eq.\eqref{eq: Omega_tt_def} can be found in Appendix A.  In the limit for $M\rightarrow0$ we obtain:


\begin{equation} \label{eq: L_tt}
	\Omega_{tt}(u_{\vec{k},\lambda},u_{\vec{k},\lambda})=-8i E_k ^2 C^{-4}(t) \left[\omega_k - \frac{3\mathcal{Y}}{16} \lambda C^2 (t) G(t) \right]\,,
\end{equation}
where $G(t)\equiv \breve{T}^i \hat{k}^i$ while $E_{\vec{k}}\equiv\frac{\omega_{k}+MC}{(2\pi)^{\frac{3}{2}}\sqrt{k^{2}+(\omega_{k}+MC)^{2}}}\rightarrow_{M\rightarrow0} \frac{1}{(2 \pi)^\frac{3}{2}} \frac{1}{\sqrt{2}}$
is obtained by the normalisation condition $\left|\Phi_{\vec{k}}\right|^{2}+\left|\gamma_{\vec{k}}\right|^{2}=\frac{1}{(2\pi)^{\frac{3}{2}}}$. In a similar way, the following relationships can be found:

\begin{align} \label{eq: Omega_tt}
	\Omega_{tt}(v_{\vec{k},\lambda},v_{\vec{k},\lambda})&=-\Omega_{tt}(u_{\vec{k},\lambda},u_{\vec{k},\lambda})\,, \notag \\
	\Omega_{tt}(u_{\vec{k},\lambda},v_{\vec{k},\overline{\lambda}})&=0 \,, \\
	\Omega_{tt}(v_{\vec{k},\lambda},u_{\vec{k},\overline{\lambda}})&=0 \,. \notag 		
\end{align}

The general form of the vacuum expectation value of the energy-momentum tensor in terms of the Bogoliubov coefficients, in the case of $\mu=0$ and $\nu=0$, is given by:
\begin{equation}
		\left\langle 0_{c}(t_0) \left| :T^{QFT}_{0 0}: \right| 0_{c}(t_0) \right\rangle=4 \frac{C^{-4}(t)}{\left(2\pi\right)^3 } \int d^3k \left\{\omega_k \left[|N_{\vec{k},+-}(t_0)|^2+|N_{\vec{k},-+}(t_0)|^2\right]+\left(-\frac{3\mathcal{Y}}{16}C^2(t)G(t)\right)\left[|N_{\vec{k},+-}(t_0)|^2-|N_{\vec{k},-+}(t_0)|^2\right]\right\}\,.
\end{equation}

Observe that the auxiliary tensor $\Omega_{\mu\nu}(A,B)$ is symmetrical by definition for each $A$ and $B$, moreover, using the Dirac equation, the trace of this tensor is zero in the case $M\rightarrow0$, as can be seen from the following relation:
\begin{equation}\label{eq: Omega_mu^mu}
	\Omega_{\mu}^{\mu}(A,B)=g^{\mu\nu}L_{\mu\nu}=2 \left(\bar{A}\tilde{\gamma}^\mu (x) D_\mu B - D_\mu \bar{A} \tilde{\gamma}^\mu (x) B\right)=-4iM\bar{A}B\,.
\end{equation}
Furthermore, it is possible to show that $\Omega_{ti}(u(v),u(v))$ is an odd function of the moment $\vec{k}$ and that $\Omega_{ij}(u(v),u(v))$ is an odd function with respect to $k_i$ and $k_j$. In particular, it is shown that $\Omega_{ti}(a,b) = k_i h_{a,b}(k)$, with $a, b = u_{\vec{k}, \lambda}, v_{\vec{k}, \lambda}$ and $h_{a,b}(k)$ a function only of the modulus of $k$. Similarly, $\Omega_{ij}(a,b) = k_i k_j s_{a,b}(k)$, with $a, b = u_{\vec{k}, \lambda}, v_{\vec{k}, \lambda}$ and $s_{a,b}(k)$ a function only of the modulus of $k$. Such properties of the auxiliary tensor $\Omega$ lead to the non-diagonal components of the vacuum expectation value of the energy-momentum tensor being zero. In particular $\left\langle 0_{c}(t_0)  \left| :T^{QFT}_{ti}: \right| 0_{c}(t_0) \right\rangle = 0$ since they are dependent on an odd function in $k_i$ integrated over the domain $k_i \in (-\infty, +\infty)$; similarly, the components $\left\langle 0_{c}(t_0) \left| :T^{QFT}_{ij}: \right| 0_{c}(t_0) \right\rangle$ turn out to be zero in the case $i \neq j$ since these quantities are dependent on a function of the form $k_i k_j n(k)$, with $n(k)$ dependent only on the modulus of $k$, integrated over a domain $(k_i, k_j) \in (-\infty, +\infty) \cross (-\infty, +\infty)$. These conclusions are not valid in the case where $i = j$ since in that case the integrand will be an even function in $k_i$. Assuming isotropy of the moments, it is found that for $i = 1, 2, 3$, the components $\left\langle 0_{c}(t_0)  \left| :T^{QFT}_{ii}: \right| 0_{c}(t_0) \right\rangle$ are identical to each other. Therefore, we conclude this analysis by saying that the expectation value on the vacuum $\left\langle 0_{c}(t_0) \right|$ of the energy-momentum tensor $T_\mu^{\nu\,\,QFT}$ is non-zero only for $\mu = \nu$. Precisely for this reason, this tensor can be interpreted as the energy-momentum tensor of a perfect fluid. Furthermore, in the massless case, the Eq.\eqref{eq: Omega_mu^mu} allows the components on the diagonal to be calculated using the following relation:

\begin{equation}
	\left\langle 0_{c}(t_0) \left| :T_{i}^{i\,\,QFT}: \right| 0_{c}(t_0) \right\rangle = \frac{\left\langle 0_{c}(t_0) \left| :T^{QFT}_{00}: \right| 0_{c}(t_0) \right\rangle}{3} \,,
\end{equation}
where no sum is intended over the index i.
Then the equation of state reads:
\begin{equation}
	w\equiv\frac{p}{\rho}=\frac{\left\langle 0_{c}(t_0) \left| :T_{i}^{i\,\,QFT} \right| 0_{c}(t_0) \right\rangle}{\left\langle 0_{c}(t_0) \left| :T^{QFT}_{00}: \right| 0_{c}(t_0) \right\rangle}=\frac{1}{3}\,.
\end{equation}
This adiabatic coefficient is typical of radiation. The importance of this index is in its impact on the dynamics of the expansion of the universe.

\subsection{Conservation laws}
We show that the covariant divergence of the expectation value of the energy-momentum tensor on the vacuum $  \left| 0_c(t_0) \right\rangle $ is related to tensor $\breve{T}^\rho$. It is well known in classical theory that the law of conservation of the energy-momentum tensor is satisfied (i.e. $\nabla_\mu T_{Classic}^{\mu\nu} =0$, where the $\nabla_\mu$ is the ordinary covariant derivative in General Relativity).

Let us consider the following lagrangian density
\begin{equation}
\tilde{\mathcal{L}}_Q=\tilde{\mathcal{L}}_{Dirac}+\tilde{\mathcal{L}}_{T}=\left[\frac{i}{2}\overline{\psi} \tilde{\gamma}^\sigma \overleftrightarrow{D}_\sigma \psi + \frac{3}{32}\mathcal{Y} \left(\overline{\phi}\gamma^5 \tilde{\gamma}^\tau \phi\right) \left(\overline{\psi}\gamma^5 \tilde{\gamma}_\tau \psi\right)\right]\,,
\end{equation}  
In the above equation $D_\sigma$ denotes the spinorial covariant derivative induced by the Levi--Civita connection. It is trivial to show that from such lagrangian density we obtain Eq.\eqref{eq: Dirac_equation}. The energy-momentum tensor $T^{\mu\nu}$ is defined through the following relationship:
\begin{equation}
	\label{eq: Tensorenergymomentum}
	T^{\mu\nu}=-u^{\mu}_A e^{\nu A}\,,
\end{equation}
where $e^\nu_A$ is the tetrad and $u^{\mu A}$ is defined by the following quantity:
\begin{equation}
	\label{eq: uconletetradi}
	u^A_\lambda= \frac{1}{\sqrt{-g}}\frac{\delta \left(\sqrt{-g} \tilde{\mathcal{L}}_Q \right)} {\delta e^\lambda_A}\,.
\end{equation}
Using the following identities:
\[
\left\{
\begin{array}{rcl}
	\frac{\delta}{\delta e^A_\mu} \left(e^B_\tau e^\tau_C \right) &=& \delta^B_A e^\mu_C + e^{\mu B} \eta_{A C}\,, \\
	\frac{\delta}{\delta e^A_\mu} \left(e^\sigma_B\right) &=& g^{\mu \sigma} \eta_{A B}\,,\\
	\frac{\delta}{\delta e^A_\mu} \left(h \right) &=& h e^\mu_A\,,
\end{array}
\right.
\]
we obtain the following relation:
\begin{equation}
	\frac{\delta}{\delta e^A_\mu} \left(\tilde{\mathcal{L}}_Q \right)=\tilde{\mathcal{L}}_Q e^\mu _A +\left\{\frac{3}{32}\mathcal{Y}\left[e^\mu_C\left(\overline{\phi}\gamma^5 \gamma_A \phi\right) \left(\overline{\psi}\gamma^5 \gamma^C \psi\right)+e^{\mu B}\left(\overline{\phi}\gamma^5 \gamma_B \phi\right) \left(\overline{\psi}\gamma^5 \gamma_A \psi\right)\right]+g^{\mu \sigma}\left[\frac{i}{2}\overline{\psi} \gamma_A \overleftrightarrow{D}_\sigma \psi\right]\right\}\,.
\end{equation}

Therefore, using the Eqs.\eqref{eq: Tensorenergymomentum}\eqref{eq: uconletetradi}, the energy-momentum tensor $T^{\mu \nu}$ is given by:
\begin{equation}
	\label{eq: tensor-e-m-torsion}
	T^{\mu \nu}=- \frac{i}{4} \left[\overline{\psi} \tilde{\gamma}^{(\mu} D^{\nu)} \psi -D^{(\mu} \overline{\psi}\tilde{\gamma}^{\nu)} \psi \right]-\frac{3}{64}\mathcal{Y} \left[\left(\overline{\phi}\gamma^5 \tilde{\gamma}^\nu \phi\right) \left(\overline{\psi}\gamma^5 \tilde{\gamma}^\mu \psi\right)+\left(\overline{\phi}\gamma^5 \tilde{\gamma}^\mu \phi\right) \left(\overline{\psi}\gamma^5 \tilde{\gamma}^\nu \psi\right)\right]\,.
\end{equation}
Note that the first member of the right-hand side of the equation is the energy-momentum tensor typical of QFT in the absence of torsion, defined by Eq.\eqref{eq:T^QFT}. Defining the axial current density by \( J^{5 \mu} \equiv \overline{\psi} \gamma^5 \tilde{\gamma}^\mu \psi \), the second member of the right-hand side of the equation is $\breve{T}^{(\mu}J^{5 \nu)}$.\\
By construction, the following condition is fulfilled:
\begin{equation}
\nabla_\mu T^{\mu\nu}=0\,.
\end{equation}
Let us consider the following quantity $	\left\langle 0_{c}(t_0) \left|\nabla_\mu T^{\mu\nu}(t) \right| 0_{c}(t_0) \right\rangle$, where we have fixed the time $t_0\neq t$, and the covariant derivative $\nabla_\mu$ acts with respect to the coordinate $x,y,z,t$ (it does not act on the fixed time $t_0$). The covariant derivative term on the contravariant tensor of rank two is generally expressed as follows:
\begin{equation}
 \nabla_\mu T^{\mu\nu}=\partial_\mu T^{\mu\nu}+ \Gamma_{\mu\sigma}^{\mu} T^{\sigma \nu}+\Gamma_{\mu\sigma}^{\nu} T^{\mu \sigma}\,,
\end{equation}
where $\Gamma_{\mu\sigma}^{\nu}$ are the coefficients of the Levi-Civita connection. It is trivial to show that $\nabla_\mu\left\langle 0_{c}(t_0) \left| T^{\mu i}(t) \right| 0_{c}(t_0) \right\rangle=0\,$ since the spatial variations of the  energy-momentum tensor are zero (i.e $\left\langle 0_{c}(t_0) \left|\partial_i T^{i i}(t) \right| 0_{c}(t_0) \right\rangle=0 $) and the only coefficients of the affine connection different from zero in the FLRW metric are:
\begin{equation}
	\Gamma_{0 j}^j=\frac{\dot{C}(t)}{C(t)} \quad \text{and} \quad \Gamma_{j j}^0=C(t) \dot{C}(t)\,,
\end{equation}
where $j = 1,2,3$ and there is no sum on the repeated indices.
In order to show that $\nabla_\mu\left\langle 0_{c}(t_0) \left| T^{\mu \nu}(t) \right| 0_{c}(t_0) \right\rangle=0\,$, we consider the following identity:
\begin{equation}
	\left\langle 0_{c}(t_0) \left| \nabla_\mu T^{\mu 0}(t) \right| 0_{c}(t_0) \right\rangle= 	\left\langle 0_{c}(t_0) \left| \partial_0 T^{0 0}(t) +\Gamma_{\mu 0}^{\mu} T^{0 0}+\sum_{j}\Gamma_{j j}^{0} T^{j j}\right| 0_{c}(t_0) \right\rangle \,.
\end{equation}
Since time $t_0$ is fixed we have $ \partial_0 \left| 0_c(t_0) \right\rangle=0$ which leads to the following relationship $\left\langle 0_{c}(t_0) \left| \partial_0 T^{0 0}(t) \right| 0_{c}(t_0) \right\rangle=\partial_0 \left\langle 0_{c}(t_0) \left| T^{0 0}(t) \right| 0_{c}(t_0) \right\rangle$ and as a consequence we
 obtain:
\begin{equation}
	\label{eq: conservation}
	\left\langle 0_{c}(t_0) \left| \nabla_\mu T^{\mu 0}(t) \right| 0_{c}(t_0) \right\rangle= \partial_0	\left\langle 0_{c}(t_0) \left|  T^{0 0}(t)\right| 0_{c}(t_0) \right\rangle + \Gamma_{\mu 0}^{\mu}	\left\langle 0_{c}(t_0) \left| T^{0 0}(t)\right| 0_{c}(t_0) \right\rangle + \sum_{j}\Gamma_{j j}^{0} \left\langle 0_{c}(t_0) \left| T^{j j}(t)\right| 0_{c}(t_0) \right\rangle=0\,.
\end{equation}
Substituting what has been obtained in Eq.\eqref{eq: tensor-e-m-torsion}, we rewrite the Eq.\eqref{eq: conservation} as follows:
\begin{equation}
	\left\langle 0_{c}(t_0) \left| \nabla_\mu T^{\mu \nu}_{QFT}(t) \right| 0_{c}(t_0) \right\rangle=-\frac{3}{64}\mathcal{Y}\left\langle 0_{c}(t_0) \left| \nabla_\mu \left(\breve{T}^{(\mu}(t)J^{5 \nu)}(t)\right) \right| 0_{c}(t_0) \right\rangle
\end{equation}
 Remembering that the tensor $L^\mu$ is defined by $L^{\mu}\equiv\left\langle 0_{c}(t_0)\left|\bar{\psi}\gamma^{5}\tilde{\gamma}^{\mu}\psi\right|0_{c}(t_0)\right\rangle $, we obtain the following expression:
 \begin{equation}
 \nabla_\mu	\left\langle 0_{c}(t_0) \left|  T^{\mu \nu}_{QFT}(t) \right| 0_{c}(t_0) \right\rangle=-\frac{3}{64}\mathcal{Y}\nabla_\mu \left(\breve{T}^{(\mu}(t)L^{ \nu)}(t)\right)
 \end{equation}

\section{Strong coupling limit}
In this section we consider the case of strong coupling which corresponds to the limit $\mathcal{Y}\rightarrow0$. Since $\mathcal{Y}=\frac{1}{1-4\zeta}$ the limit coincides to the limit ${\zeta\rightarrow\infty}$. Such regime allows for analytical treatmeant of the integral in Eq.\eqref{eq:vecL}.

Expanding in series the Bogoliubov coefficients within the Eq.\eqref{eq:vecL} for $\left(\frac{-3\mathcal{Y}}{8}\right)\rightarrow0$, we obtain at the lowest non-zero order in the $\mathcal{Y}$ parameter the following expression:
\[
\left(\left|N_{\vec{k},+-}\right|^{2}-\left|N_{\vec{k},-+}\right|^{2}\right)\simeq  \left(\frac{3\mathcal{Y}}{8}\right)^{3}\left(\frac{g\,|\mu_{\vec{k}}|^{2}}{k^{3}}\right)\,.
\]
In the eq.(\ref{eq:vecL}), the term $-\int d^{3}k\left[\frac{C^{-4}(t)}{(2\pi)^{3}}\left(\left(\frac{3\mathcal{Y}}{8}\right)^{3}\left(\frac{g\,|\mu_{\vec{k}}|^{2}}{k^{3}}\right)\right)\right]\simeq O\left(\mathcal{Y}^3\right)$
is negligible since it is of the third order in the parameter  $\left(\mathcal{Y}\right)$. \\
Therefore $\vec{L}$   reduces to the term containing the products
mixed products of  $N_{\vec{k}}$ and $M_{\vec{k}}$:
\begin{equation}
	\vec{L}\equiv\int d^{3}k\left\{ \frac{C^{-4}(t)}{(2\pi)^{3}}\left[\left(\vec{\varepsilon}(\hat{k})\right)\left(e^{+2i\omega_{k}\tau}\right)N_{\vec{k},+-}^{*}(t_0)M_{\vec{k},++}(t_0)+\left(\vec{\varepsilon}(\hat{k})\right)^{*}\left(e^{+2i\omega_{k}\tau}\right)N_{\vec{k},-+}(t_0)M_{\vec{k},--}(t_0)+c.c.\right]\right\} \,.
\end{equation}

Using the Mathematica programme to calculate and expand in Taylor's series the following quantity: $S_{1}\equiv\left(e^{+2i\omega_{k}\tau}\right)N_{\vec{k},+-}^{*}M_{\vec{k},++}$
and $S_{2}\equiv\left(e^{+2i\omega_{k}\tau}\right)N_{\vec{k},-+}M_{\vec{k},--}$,
we obtain for $S_{1}$:

\begin{equation}
	S_{1}=\left(\frac{-3\mathcal{Y}}{16}\right)\frac{\mu_{\vec{k}}^*}{k}+O\left(\left(\frac{-3\mathcal{Y}}{16}\right)^{2}\right)\,.
\end{equation}
and similarly for $S_{2}$:
\[
S_{2}=\left(\frac{-3\mathcal{Y}}{16}\right)\frac{\mu_{\vec{k}}}{k}+O\left(\left(\frac{-3\mathcal{Y}}{16}\right)^{2}\right)\,.
\]
Collecting the above results, one has
\begin{equation}
	\vec{L}=-\frac{3\mathcal{Y}}{8 (2\pi)^{3}}C^{-4}(t)\int d^{3}k\left(\frac{\vec{\varepsilon}(\hat{k})\mu_{\vec{k}}^{*}(t_0)+c.c}{k}\right)\,.
\end{equation}
The angular integrals are straightforward and give
\begin{equation}
	\int_{-1}^{+1}d\left(\cos\theta\right)\int_{0}^{2\pi}d\phi\left[Re\left\{ \vec{\varepsilon}(\hat{k})\mu_{\vec{k}}^{*}(t_0)\right\} \right]=\left\{ \begin{array}{c}
		\frac{8}{3}\pi\breve{T}^{1}(t_0)\,,\\
		\frac{8}{3}\pi\breve{T}^{2}(t_0)\,,\\
		\frac{8}{3}\pi\breve{T}^{3}(t_0)\,,
	\end{array}\right.
\end{equation}
so that denoting by $\Lambda$ the UV momentum cutoff, the non vanishing components of the expectation values read:
\begin{equation}
	L^{i}=-2\pi\mathcal{Y}\frac{C^{-4}(t)}{(2\pi)^{3}}\breve{T}^{i}(t_0)\int_{0}^{\Lambda}dk\,k\,.
\end{equation}
\\
Therefore, the expectation value of the spin operator $\vec{L}$ on the
vacuum $|0_{c}(t)\rangle $ is different from zero and is
\begin{equation}
	\vec{L}=\frac{1}{C(t)}\left\langle 0_{c}(t_0)\left|\bar{\psi}\gamma^{5}\vec{\gamma}\psi\right|0_{c}(t_0)\right\rangle =-\pi\mathcal{Y}\frac{C^{-4}(t)}{(2\pi)^{3}}\Lambda^{2}\left(\breve{T}^{1}(t_0),\breve{T}^{2}(t_0),\breve{T}^{3}(t_0)\right)\,.
\end{equation}

\subsection{Strong coupling limit for the energy-momentum tensor}

Now we can also compute explicitly the expectation value on the vacuum $|0_{c}(t)\rangle $ for the energy-momentum tensor. By replacing the results of Eqs.\eqref{eq: L_tt}\eqref{eq: Omega_tt} in Eq.\eqref{eq: normal_ordering_T_munu}, the following relation is obtained:

\begin{equation}
	\left\langle 0_{c}(t_0) \left| :T^{QFT}_{00}: \right| 0_{c}(t_0) \right\rangle = 4 \int d^3k \left[E^2_k C^{-4}(t) \omega_k \right] \left\{2 \left[|N_{\vec{k},+-}(t_0)|^2+|N_{\vec{k},-+}(t_0)|^2 \right] \right\}\,.
\end{equation}
The quantity in brackets was calculated using the programme Mathematica. Expanding in series the above quantity for $\left(\mathcal{Y}\right)\rightarrow0$, we arrest the series expansion in the lowest non-zero order in the parameter $\mathcal{Y}$ thereby obtaining the following quantity:
\begin{equation}
2\left(\left|N_{\vec{k},+-}\right|^{2}+\left|N_{\vec{k},-+}\right|^{2}\right)\simeq\left(\frac{3\mathcal{Y}}{8}\right)^{2}\left(\frac{|\mu_{\vec{k}}|^{2}}{k^{2}}\right)\,,
\end{equation}
where $\mu_{\vec{k}}$ is given by Eq.\eqref{eq: mu}. Collecting the above results, one has
\begin{equation}
		\left\langle 0_{c}(t_0) \left| :T^{QFT}_{00}: \right| 0_{c}(t_0) \right\rangle = 4 E_k^2 C^{-4}(t) \left( \frac{3\mathcal{Y}}{8} \right)^2 \int d^3k \left(\frac{|\mu_{\vec{k}}(t_0)|^{2}}{k^{2}}  \right)\omega_k\,,
\end{equation}
where $\omega_k$ in the limit for $M\rightarrow0$ tends to $k$. The angular integrals are straightforward and give:
\begin{equation}
	\int_{-1}^{+1}d\left(\cos\theta\right)\int_{0}^{2\pi}d\phi\,\,|\mu_{\vec{k}}(t_0)|^{2} = \frac{8}{3} \pi \left[\left(\breve{T}^1(t_0)\right)^2 +\left(\breve{T}^2(t_0)\right)^2+\left(\breve{T}^3 (t_0) \right)^2 \right]
\end{equation}
so that denoting by $\mathcal{K}$ the UV momentum cutoff, the expectation values read:
\begin{equation}
	\left\langle 0_{c}(t_0) \left| :T^{QFT}_{00}: \right| 0_{c}(t_0) \right\rangle = \frac{C^{-4}(t)}{3 \pi^2} \left( \frac{3\mathcal{Y}}{8} \right)^2 \left[\left(\breve{T}^1(t_0)\right)^2 +\left(\breve{T}^2(t_0)\right)^2+\left(\breve{T}^3(t_0)\right)^2 \right] \left(\mathcal{K}\right)^2
\end{equation}
\\
Therefore, the expectation value of the $tt$ components of the tensor energy-momemtun $T_{\mu\nu}$ on the
vacuum $|0_{c}(t)\rangle $ is different from zero and using Eq.\eqref{eq: componenti} it is
\begin{equation}
	\begin{aligned}
		\left\langle 0_{c}(t_0) \left| :T^{QFT}_{00}: \right| 0_{c}(t_0) \right\rangle &= \frac{C^{-4}(t)C^{-6}(t_0)}{3 \pi^2} \left( \frac{3\mathcal{Y}}{8} \right)^2 \left(\mathcal{K}\right)^2 \Bigg[Z^2 + 4 \left(|H_2|^4 + |H_3|^4\right)+ \\
		&\quad + 4 |H_3|^2 \left(Z + |H_4|^2\right) + 4 \left(-Z |H_2|^2 + |H_1|^2|H_2|^2\right)+ \\
		&\quad + 4 \left(-2|H_2|^2|H_3|^2\right) + 4 H_1 H_2^* H_3^* H_4  + 4 H_1^* H_2 H_3 H_4^* \Bigg] \, .
	\end{aligned}
\end{equation}
where $K$ is fixed by Eq.\eqref{eq: psibarrapsi}. Note that the vacuum expetation value of the energy-momentum tensor  $\left\langle 0_{c}(t_0) \left| :T_{00}: \right| 0_{c}(t_0) \right\rangle=\left\langle 0_{c}(t_0) \left| :T^{QFT}_{00}: \right| 0_{c}(t_0) \right\rangle$ since for $\mu=0$ the tensor $\breve{T}^\mu$ is zero.

\section{Discussions and conclusions}

We considered a classical Dirac field $\phi$ in a square-torsion theory, acting as a source of torsion for a second Dirac field $\psi$ in an FLRW background. We have shown that the non diagonal Hamiltonian of $\psi$ is diagonalized by means of a Bogoliubov transformation leading to a new Fock space representation, which turns out to be unitarily inequivalent to that of $\psi$ in the absence of torsion.
Therefore, the transformed vacuum $|0_c(t_0)\rangle$ also turns out to be orthogonal  to the original vacuum $|0\rangle$ in the infinite volume limit and has a condensate structure  which presents   non-trivial contributions to the  energy-momentum  tensor and to the axial current, which modify the dynamics of the classical field $\phi$ and of the field equations, as back-reaction.
In order to take into account these contributions, we considered a new classical field with an associated torsion term, which generated further contributions to the vacuum expectation values of both the energy-momentum tensor and the axial current, affecting the field equations through an iterative process.

In the first step of the iterative process, we obtained a contribution to the energy and to the torsion proportional to $C^{-4}(t)$, in the strong coupling limit. This result, provides indications that the condensate structure of the vacuum could affect the inflation in the early stage of the universe. Other steps of the iterative process indicate that the vacuum condensate potentially could affect dark energy and dark matter. 
A deep analysis of the full back-reaction mechanism will be performed in future work. Similar result may be expected even in absence of the classical fermion $\phi$, as a condensation effect induced by axial fermion self interaction due to the torsion \cite{Quaranta-Capolupo}.

\section*{Appendix A: The auxiliary tensor}
By definition, the auxiliary tensor $\Omega_{\mu\nu}(A,B)$ is defined as follows:
\begin{equation*}
	\Omega_{\mu\nu}(A,B)=\overline{A}\tilde{\gamma}_\mu D_\nu B+\overline{A}\tilde{\gamma}_\nu D_\mu B - D_\mu \overline{A} \tilde{\gamma}_\nu B- D_\nu \overline{A} \tilde{\gamma}_\mu B \, . \notag
\end{equation*}
The calculation of $\Omega_{\mu\nu}(A,B)$ for $\mu=0$, $\nu=0$, $A=B=u_{\vec{k},\lambda}$ is explicitly shown below.
\begin{align*}
	\Omega_{tt}(u_{\vec{k},\lambda},u_{\vec{k},\lambda}) &\equiv 2 \left[\overline{u}_{\vec{k},\lambda} \tilde{\gamma}_t D_t u_{\vec{k},\lambda} - \left( D_t \overline{u}_{\vec{k},\lambda} \right) \tilde{\gamma}_t u_{\vec{k},\lambda} \right]= \notag \\
	&= 2 \left[u_{\vec{k},\lambda}^\dagger \partial_t u_{\vec{k},\lambda}-\left(\partial_t u_{\vec{k},\lambda}^\dagger\right) u_{\vec{k},\lambda} \right]\, ,
\end{align*}

where the relations $\tilde{\gamma}_t=\gamma_0$ and $D_t u_{\vec{k},\lambda}=\partial_t u_{\vec{k},\lambda}$ were used. Using the ansatz $u_{\vec{k},\lambda}(t,\boldsymbol{x})=e^{i\boldsymbol{k}\cdot\boldsymbol{x}}\left(\begin{array}{c}
	f_{k,\lambda}(t)\xi_{\lambda}(\hat{k})\\
	g_{k,\lambda}(t)\lambda\xi_{\lambda}(\hat{k})
\end{array}\right)$ for positive energies the following quantity is obtained:


\begin{align*}
	\Omega_{tt}(u_{\vec{k},\lambda}, u_{\vec{k},\lambda}) &= 2 \left[\left( f_{k,\lambda}^*(t)\xi^\dagger_{\lambda}(\hat{k}), \, g_{k,\lambda}^*(t)\lambda\xi^\dagger_{\lambda}(\hat{k}) \right) \left(\begin{array}{c}
		\left( \partial_t f_{k,\lambda}(t) \right)\xi_{\lambda}(\hat{k})\\
		\left( \partial_t g_{k,\lambda}(t) \right) \lambda\xi_{\lambda}(\hat{k})
	\end{array}\right) + \right. \notag \\
	&\left. \quad - \left( \partial_t \left( f^*_{k,\lambda}(t) \right) \xi^\dagger_{\lambda}(\hat{k}), \, \partial_t \left( g^*_{k,\lambda}(t) \right) \lambda\xi^\dagger_{\lambda}(\hat{k}) \right)\left(\begin{array}{c}
		f_{k,\lambda}(t)\xi_{\lambda}(\hat{k})\\
		g_{k,\lambda}(t)\lambda\xi_{\lambda}(\hat{k})
	\end{array}\right) \right] \,.
\end{align*}

We rescale the functions as follows:
\begin{equation*}
	\Phi_{\vec{k},\lambda}\equiv C^{\frac{3}{2}}f_{\vec{k}, \lambda \,,}\qquad\gamma_{\vec{k},\lambda}\equiv C^{\frac{3}{2}}g_{\vec{k},\lambda}\,,
\end{equation*}
so that the previous equation became:

\begin{align*}
	\Omega_{tt}(u_{\vec{k},\lambda}, u_{\vec{k},\lambda}) &= 2 C^{-3}(t) \left[\left(\Phi_{\vec{k},\lambda}^*(t)\xi^\dagger_{\lambda}(\hat{k}), \, \gamma_{\vec{k},\lambda}^*(t)\lambda\xi^\dagger_{\lambda}(\hat{k}) \right) \left(\begin{array}{c}
		\left( \partial_t \Phi_{\vec{k},\lambda}(t) \right)\xi_{\lambda}(\hat{k})\\
		\left( \partial_t \gamma_{\vec{k},\lambda}(t) \right) \lambda\xi_{\lambda}(\hat{k})
	\end{array}\right) + \right. \notag \\
	&\left. \quad - \left( \partial_t \left( \Phi^*_{\vec{k},\lambda}(t) \right) \xi^\dagger_{\lambda}(\hat{k}), \, \partial_t \left( \gamma^*_{k,\lambda}(t) \right) \lambda\xi^\dagger_{\lambda}(\hat{k}) \right)\left(\begin{array}{c}
		\Phi_{\vec{k},\lambda}(t)\xi_{\lambda}(\hat{k})\\
		\gamma_{k,\lambda}(t)\lambda\xi_{\lambda}(\hat{k})
	\end{array}\right) \right] \,.
\end{align*}

Using the property $\xi^\dagger_{\lambda}(\hat{k})\xi_{\lambda}(\hat{k})=1$, we obtain the result of Eq.\eqref{eq: Omega_tt_def}:
\begin{align*}
	\Omega_{tt}(u_{\vec{k},\lambda},u_{\vec{k},\lambda}) &= 2 C^{-3}(t) \xi^\dagger_{\lambda}(\hat{k})\xi_{\lambda}(\hat{k}) \left[ \Phi^*_{\vec{k},\lambda}(t) {\partial_t} \Phi_{\vec{k},\lambda}(t) +  \gamma^*_{\vec{k},\lambda}(t) {\partial_t} \gamma_{\vec{k},\lambda}(t)  \right]+ \notag \\
	& \quad \quad - \left( \partial_t \Phi^*_{\vec{k},\lambda}(t) \right) \Phi_{\vec{k},\lambda}(t) - \left( \partial_t \gamma^*_{\vec{k},\lambda}(t) \right) \gamma_{\vec{k},\lambda}(t) =\notag \\
	& \equiv 2 C^{-3}(t) \left[ \Phi^*_{\vec{k},\lambda}(t) \overleftrightarrow{\partial_t} \Phi_{\vec{k},\lambda}(t) + \gamma^*_{\vec{k},\lambda}(t) \overleftrightarrow{\partial_t} \gamma_{\vec{k},\lambda}(t)\right]\,.
\end{align*}


\begin{thebibliography}{99}






\bibitem{Ligo0}
B. P. Abbott, R. Abbott, T. D. Abbott, M. R. Abernathy, F. Acernese, K. Ackley, C. Adams, T. Adams, P. Addesso \textit{et al.}, \textit{Phys. Rev. Lett.} \textbf{116}, 061102 (2016).

\bibitem{Ligo1}
The LIGO Scientific Collaboration, J Aasi, B P Abbott, R Abbott, \textit{et al.}, \textit{Class. Quantum Grav.}, \textbf{32}, 074001 (2015).

\bibitem{Virgo0}
F Acernese \textit{et al.}, \textit{Class. Quantum Grav.}, \textbf{32}, 024001 (2015).

\bibitem{Virgo1}
B. P. Abbott, R. Abbott, T. D. Abbott, F. Acernese, K. Ackley, C. Adams, T. Adams, P. Addesso, R. X. Adhikari \textit{et al.} (LIGO Scientific Collaboration and Virgo Collaboration), \textit{Phys. Rev. Lett.} \textbf{119}, 161101 (2017).

\bibitem{Virgo2}
B. P. Abbott, R. Abbott, T. D. Abbott, S. Abraham, F. Acernese, K. Ackley, C. Adams, R. X. Adhikari, V. B. Adya \textit{et al.} (LIGO Scientific Collaboration and Virgo Collaboration), \textit{Phys. Rev. X} \textbf{9}, 031040 (2019).

\bibitem{Rubin1}
V. Rubin and W. K. Ford, 
\textit{Astrophys. J.} \textbf{159}, 379-403 (1970).

\bibitem{Rubin2}
V. Rubin and W. K. Ford, 
\textbf{14}, 479-539 (1976).

\bibitem{Rubin3}
V. Rubin and J. D. P. Kenney, 
\textit{Astron. J.} \textbf{108}, 1943-1956 (1994).

\bibitem{Rubin4}
V. Rubin and M. Merrifield, 
 \textit{Phys. Today} \textbf{53}, 4, 41–47 (2000).

\bibitem{Rubin5}
V. Rubin and W. K. Ford, 
\textit{Sci. Amer.} \textbf{242}, 96-103 (1980).

\bibitem{PecceiQuinn77}
R. D. Peccei and H. R. Quinn, 
 \textit{Phys. Rev. Lett.} \textbf{38}, 1440-1443 (1977).

\bibitem{DMAssion0}
S. Weinberg, 
 \textit{Phys. Rev. D} \textbf{19}, 1277–1289 (1978).

\bibitem{DMAssion1}
M. Dine, W. Fischler, and M. Srednicki,
\textit{Phys. Lett. B} \textbf{104}, 199-202 (1981).

\bibitem{DMAssion2}
J. Preskill, M. B. Wise, and F. Wilczek, 
\textit{Phys. Lett. B} \textbf{120}, 127-132 (1983).

\bibitem{DMAssion3}
ADMX Collaboration,
 \textit{Phys. Rev. Lett.} \textbf{123}, 161301 (2019).

\bibitem{DMAssion4}
G. Rybka et al., 
\textit{Phys. Rev. D} \textbf{89}, 021101 (2014).

\bibitem{DMAssion5}
D. J. E. Marsh, 
\textit{Phys. Rep.} \textbf{643}, 1–79 (2016).

\bibitem{Axion4}
A. Capolupo, I. De Martino, G. Lambiase and An. Stabile, \textit{Phys. Lett. B} \textbf{790}, pp. 427--435 (2019).

\bibitem{HP1969}
S. W. Hawking and R. Penrose, 
\textit{Proc. Roy. Soc. Lond. A} \textbf{314}, 529-548 (1969).

\bibitem{DerivativeOrder0}
S. Nojiri, S. D. Odintsov, and V. K. Oikonomou, 
 \textit{Phys. Rept.} \textbf{692}, 1-93 (2017).

\bibitem{Capozziello2011}
S. Capozziello and M. De Laurentis, 
 \textbf{509}, 167-321 (2011).

\bibitem{DerivativeOrder1}
S. A. Appleby and R. A. Battye,
\textit{Phys. Rev. D} \textbf{76}, 103510 (2007).

\bibitem{DerivativeOrder2}
S. Nojiri and S. D. Odintsov, 
\textit{Phys. Rev. D} \textbf{68}, 123512 (2003).

\bibitem{DerivativeOrder3}
A. A. Starobinsky, 
 \textit{J. High Energy Phys.} \textbf{2007}, 1-32 (2007).
 





\bibitem{Cai:2013lqa}
Yi-Fu Cai, F.Duplessis, E.N.Saridakis, 
\textit{Phys. Rev. D} \textbf{90}, 064051 (2014).



\bibitem{Nashed:2020kdb}
G.G.L.Nashed, E.N.Saridakis, 
\textit{Phys. Rev. D} \textbf{102}, 124072 (2020).

\bibitem{Capozziello:2011gw}
S.Capozziello, S.Vignolo, 
\textit{Int. J. Geom. Meth. Mod. Phys.} \textbf{9},1250006(2012).

\bibitem{Vignolo:2018eco}
S.Vignolo, R.Cianci, S.Carloni, 
\textit{Class. Quant. Grav.} \textbf{35},095014(2018).

\bibitem{Vignolo:2019qcg}
S.Vignolo, 
\textit{Universe} \textbf{5}, 224(2019).


\bibitem{Torsion6} 
S. Capozziello, M. De Laurentis, L. Fabbri and S. Vignolo, \textit{Eur. Phys. J. C} \textbf{72}, 1908 (2012).
\bibitem{DETorsion} 
A. van de Venn, D. Vasak, J. Kirsch, J. Struckmeier, \textit{Eur. Phys. J. C} \textbf{83}, 288 (2023).
\bibitem{Cirillo-Lombardo} 
D. J. Cirilo-Lombardo, \textit{EPL} \textbf{127}, 10002 (2019).

\bibitem{FV1} 
L. Fabbri and S. Vignolo, \textit{Class. Quantum Grav.} \textbf{28}, 125002 (2011).

\bibitem{VFC} S. Vignolo, L. Fabbri and R. Cianci, \textit{J. Math. Phys.} \textbf{52}, 112502 (2011).

\bibitem{Torsion1}

N. E. Mavromatos, P. Pais and A. Iorio, \textit{Universe} \textbf{2023}, 9 (12), 516 (2023).

\bibitem{Hehl1976}
F. W. Hehl, P. von der Heyde, G. D. Kerlick and
J. M. Nester, \textit{Rev. Mod. Phys.} \textbf{48}, 3 (1976).

\bibitem{Torsion2}
M. F. Ciappina, A. Iorio, P. Pais and A. Zampeli, \textit{Phys. Rev. D} \textbf{101}, 036021 (2020).


\bibitem{Torsion5}
L. Fabbri and S. Vignolo,  \textit{Class. Quantum Grav.} \textbf{28}, 125002 (2011).


\bibitem{Torsion4}
S. Vignolo, S. Carloni and L. Fabbri, \textit{Phys. Rev. D} \textbf{91}, 043528 (2015).


 \bibitem{Kibble1967}
T. W. B. Kibble, 
\textit{Phys. Lett.} \textbf{12}, 138-141 (1967).

\bibitem{Kibble1976}
T. W. B. Kibble, 
\textbf{9}, 1387-1398 (1976).

\bibitem{Kibble1982}
T. W. B. Kibble, 
\textit{Nature} \textbf{302}, 508-510 (1982).

\bibitem{KibbleZurek1983}
T. W. B. Kibble and W. H. Zurek, 
\textit{Nature} \textbf{305}, 636-638 (1983).
\bibitem{Kibble1980}
T. W. B. Kibble, 
\textit{Phys. Rep.} \textbf{67}, 183-287 (1980).








\bibitem{S}
D. W. Sciama, 
\textit{Rev. Mod. Phys.} \textbf{36}, 463 (1964).

\bibitem{K}
T. W. B. Kibble, 
\textit{J.Math.Phys.} \textbf{2}, 212 (1961).

\bibitem{k}
G. D. Kerlick, 
\textit{Phys. Rev. D} \textbf{12}, 3004 (1975).

\bibitem{Inomata:1976wi}
A. Inomata, 
\textit{Phys.Rev.D} \textbf{18}, 3552 (1978).



\bibitem{Magueijo:2012ug}
J.Magueijo, T.G.Zlosnik, T.W.B.Kibble, 
\textit{Phys.Rev.D} \textbf{87}, 063504 (2013).

\bibitem{Khriplovich:2013tqa}
I.B.Khriplovich, A.S.Rudenko, 
 \textit{JHEP} \textbf{1311}, 174 (2013).
\bibitem{Alexander:2014eva}
S.Alexander, C.Bambi, A.Marciano, L.Modesto, 
\textit{Phys.Rev.D} \textbf{90}, 123510 (2014).
\bibitem{Buchbinder:1985ym}
I.L.Buchbinder, S.D.Odintsov, I.L.Shapiro, \textit{Phys.Lett.B} \textbf{162}, 92 (1985).
\bibitem{Fabbri:2012yg}
Luca Fabbri, Stefano Vignolo, 
\textit{Int. J. Theor. Phys.} \textbf{51}, 3186 (2012).



\bibitem{VFS}
S. Vignolo, L. Fabbri and C. Stornaiolo, \textit{Ann. Phys. (Berlin)} \textbf{524}, 826-839 (2012).

\bibitem{Shapiro} I. L. Shapiro, Physics Reports \textbf{357}, pp. 113-213 (2002).

\bibitem{QCD}
B. L. Ioffe, \textit{Physics of Atomic Nuclei} \textbf{66}, pp. 30-43 (2003).

\bibitem{QCD2}
Y. Nambu and G. Jona-Lasinio, \textit{Phys. Rev} \textbf{122}, pp. 345-358 (1961).

\bibitem{Hawking}
S. W. Hawking, \textit{Commun. Math. Phys.} \textbf{43}, pp. 199-220 (1975)

\bibitem{Unruh}
W. G. Unruh, \textit{Phys. Rev. D} \textbf{14} (4), 870-892 (1976).

\bibitem{Takagi}
S. Takagi, \textit{Progress of Theoretical Physics Supplement} \textbf{88}, pp. 1-142 (1986).

\bibitem{Vanzella}
D. A. T. Vanzella and G. E. Matsas, \textit{Phys. Rev. Lett.} \textbf{87}, 151301 (2001).

\bibitem{Parker}
L. Parker, \textit{J. Phys. A: Math. Theor.} \textbf{45}, 374023 (2012).

\bibitem{Parker2}
N. Birrell and P. Davies, 
\textit{Cambridge University Press}, England, 1982.

\bibitem{Casimir}
J. C. Da Silva, F. C. Khanna, A. Matos Neto and A. E. Santana, \textit{Phys. Rev. A} \textbf{66}, 052101 (2002).

\bibitem{Lepto1} W. Buchmüller and M. Plümacher, \textit{Phys. Lett. B} \textbf{511}, 74-76 (2001).


\bibitem{Kaplan} D. B. Kaplan, A. E. Nelson and N. Weiner, Phys. Rev. Lett. \textbf{93}
, 091801 (2004).

\bibitem{N1} S. M. Bilenky and B. Pontecorvo, Phys. Rep. \textbf{41},
225 (1978).

\bibitem{N2} S. M. Bilenky and S. T. Petcov, Rev. Mod. Phys. \textbf{59},
671 (1987).

\bibitem{N3} G. Aad et al. (ATLAS Collaboration), Phys. Lett. B \textbf{716.1}
(2012).

\bibitem{N4} Y. Fukuda et al., Phys. Rev. Lett. \textbf{81} (8 Aug.
1998), pp. 1562-1567.

\bibitem{N5-0} M. Blasone, G. Vitiello, Annals Phys. \textbf{244},
283-311 (1995);

\bibitem{N5}
M. Blasone, A. Capolupo, G. Vitiello, Phys. Rev. D
\textbf{66}, 025033 (2002);

\bibitem{N5-1} A. Capolupo, S. Capozziello, G. Vitiello, Phys. Lett.
A \textbf{363}, 53, (2007); A. Capolupo, Adv. in High En. Phys., \textbf{Volume 2016}, 8089142, 10 pages (2016);
A. Capolupo,  Adv. in High Ene. Phys., \textbf{Volume 2018}, 9840351, 7 pages (2018).

\bibitem{N6} K. Fujii, C. Habe and T. Yabuki, Phys. Rev. D \textbf{59},
113003 (1999).

\bibitem{N7} K.C. Hannabuss and D.C. Latimer, J. Phys. A \textbf{33},
1369 (2000).

\bibitem{N8} M. Blasone, A. Capolupo, O. Romei, G. Vitiello, Phys.
Rev. D \textbf{63}, 125015, (2001).

\bibitem{N9} E. Alfinito, M.Blasone, A.Iorio, G.Vitiello, Phys. Lett.
B \textbf{362}, 91 (1995).

\bibitem{N10} Y. Grossman and H. J. Lipkin, Phys. Rev. D \textbf{55},
2760 (1997).

\bibitem{N11} D. Piriz, M. Roy and J. Wudka, Phys. Rev. D \textbf{54},
1587 (1996).

\bibitem{N12} C. Y. Cardall and G. M. Fuller, Phys. Rev. D \textbf{55},
7960 (1997).

\bibitem{N13} L. Buoninfante, G. G. Luciano, L. Petruzziello and
L. Smaldone, Phys. Rev. D \textbf{101}, 024016 (2020).

\bibitem{N14} A. Capolupo, S. M. Giampaolo and A. Quaranta, Phys.
Lett. B \textbf{820}, 136489 (2021).

\bibitem{N15} G. G. Luciano, EPJ Plus \textbf{138}, 83 (2023).

\bibitem{N16} A. Capolupo and A. Quaranta, J. Phys. G \textbf{50},
055003 (2023).

\bibitem{N17} A. Capolupo and A. Quaranta, Phys. Lett. B \textbf{839},
137776 (2023).

\bibitem{Torsioncostant}
A. Capolupo, G. De Maria, S. Monda, A. Quaranta, R. Serao, 10, (4), 170  \textit{Universe} \textbf{2024} (2024).

\bibitem{Neut2}
A. Capolupo, S. Carloni and A. Quaranta, \textit{Phys. Rev. D} \textbf{105}, 105013 (2022).


\bibitem{Neut4}
A. Capolupo, A. Quaranta and R. Serao, \textit{Symmetry} \textbf{2023}, 15(4), 807 (2023).

\bibitem{Neut5}
A. Capolupo and A. Quaranta, \textit{Phys. Lett. B} \textbf{840}, 137889 (2023).

\bibitem{Neut6}
A. Capolupo, S. M. Giampaolo, G. Lambiase and A. Quaranta, \textit{Eur. Phys. J. C} \textbf{80}, 423 (2020).
\bibitem{K4} C. R. Ji and Y. Mishchenko, \textit{Phys. Rev. D} \textbf{65}, 096015 (2002).
\bibitem{K2} K. Fujii, C. Habe and T. Yabuki, \textit{Phys. Rev. D} \textbf{64}, 013011 (2001).
\bibitem{K5} C. R. Ji and Y. Mishchenko, \textit{Ann. Phys.} \textbf{315}, Issue 2, pp. 488-504 (2005).
\bibitem{Barker1978} B. M. Barker and R. F. O'Connell, \textit{Gen. Rel. Grav.} \textbf{11}, 2 (1979).
\bibitem{NTorsion7} M. Adak, T. Dereli, H. Ryder, \textit{Class. Quantum Grav.} \textbf{18}, 1503-1512 (2001).
\bibitem{SUGRA} P. Van Nieuwenhuizen, \textit{Phys. Rept.} \textbf{68}, 4, pp. 189-398 (1981).
\bibitem{Jenkins} A. Jenkins, \textit{Phys. Rev. D} \textbf{69}, 105007 (2004).
\bibitem{Sakharov} A. D. Sakharov, \textit{Sov. Phys. Usp.} \textbf{34}, 392 (1991).


\bibitem{Pisacane}
A. Capolupo, S. Capozziello, G. Pisacane, and A. Quaranta, 
\textit{Phys. Dark Univ.} \textbf{48}, 101894 (2025).




 \bibitem{CapDark1}
A. Capolupo, \textit{Adv. High En. Phys.} {\bf 2018}, 9840351 (2018).

 \bibitem{CapDark2}
A. Capolupo, \textit{Adv. High En. Phys.} {\bf 2016}, 8089142 (2016).

\bibitem{CapDark3}
A. Capolupo, S. Capozziello, G. Vitiello, \textit{Phys. Lett. A} \textbf{373}, pp. 601--610 (2009).

\bibitem{CapDark4}
A. Capolupo, S. Capozziello, G. Vitiello, \textit{Phys. Lett. A} \textbf{363}, 53 (2007).

\bibitem{CapDark5}
M. Blasone, A. Capolupo, S. Capozziello, S. Carloni, G. Vitiello, Phys. Lett. A \textbf{323}, pp. 182--189 (2004).

\bibitem{GPL}
G.~G.~Luciano,
Eur. Phys. J. C \textbf{83} no.4, 329 (2023)

G.~G.~Luciano and J.~Gin\'e,
Phys. Dark Univ. \textbf{41}, 101256 (2023)

G.~G.~Luciano,
Phys. Rev. D \textbf{106} no.8, 8 (2022)




\bibitem{Cabral} F. Cabral, F. Lobo and D. Rubiera-Garcia,Class.
Quantum Grav. \textbf{38}, 195008 (2021).

\bibitem{Cap2}
A. Capolupo, C.R. Ji, Y. Mischenko, G. Vitiello, 
\textit{Phys. Lett. B} \textbf{594}, 135-140, (2004).

\bibitem{Capoz1}
S. Capozziello, M. De Laurentis, Physics Reports \textbf{509}, 4-5, pp. 167-321 (2011).

\bibitem{Propagating1}
D. Benisty, E. I. Gundelman, A. Van de Venn, D. Vasak, J. Struckheimer and H. Stoecker, \textit{Eur. Phys. J. C} \textbf{82}, 264 (2022).

\bibitem{Propagating2}
S. M. Carrol and G. B. Field, \textit{Phys. Rev. D} \textbf{50}, 3867 (1994).

\bibitem{Propagating3}
Y.-F. Cai, S. Capozziello, M. De Laurentis and E. N. Saridakis, \textit{Rep. Prog. Phys.} \textbf{79}, 106901 (2016).

\bibitem{Cosm1} G. Tukhashvili and P. J. Steinhardt, \textit{Phys. Rev. Lett.} \textbf{131}, 091001 (2023).
\bibitem{LSB} D. Colladay and V. Alan Kosteleck\'{y}, \textit{Phys. Rev. D} \textbf{58}, 116002 (1998).

\bibitem{LSB2} V. Alan Kosteleck\'{y} and R. Lehnert, \textit{Phys. Rev. D} \textbf{63}, 065008 (2001).

\bibitem{Campbell} J. E. Campbell, Proceedings of the London Mathematical Society \textbf{28}, 381–390 (1897); Proceedings of the London Mathematical Society \textbf{29}, 14–32 (1898).
\bibitem{Campbell2} H. F. Baker, Proceedings of the London Mathematical Society (1) \textbf{34}, 347–360 (1902); Proceedings of the London Mathematical Society (1) \textbf{35}, (1903) 333–374; Proceedings of the London Mathematical Society (Ser 2) \textbf{3}, 24–47 (1905).
\bibitem{Campbell3} F. Hausdorff, 
Ber Verh Saechs Akad Wiss Leipzig \textbf{58}, 19–48 (1906).
\bibitem{Campbell4} A. Perelomov, \textit{Generalized coherent states and their applications}, Springer, Berlin 1986.


\bibitem{Umezawa1}
H. Umezawa, H. Matsumoto and M. Tachiki, 
North-Holland Publishing Company, 1982, ISBN 978-0444863614.

\bibitem{Umezawa2}
H. Umezawa, 
 American Institute of Physics, 1993, ISBN 978-1563960819.




\bibitem{Quaranta-Capolupo}
A. Capolupo, A. Quaranta \textit{J. Phys. G} \textbf{51}, 105202 (2024).

\end{thebibliography}
\end{document}